\documentclass[pre,twocolumn,showpacs,amsmath,amssymb,floatfix,eqsecnum,superscriptaddress]{revtex4}

\usepackage{graphicx}
\usepackage{times}

\begin{document}

\newcommand\QOK[1]{{\bf QOK: #1}}
\newcommand\query[1]{{\bf Q-SS: #1}}
\newcommand\todo[1]{{\bf CLH TODO:#1}}
\newcommand{\SAVE}[1]{{}}
\def\OMIT#1 {{}}   

\def \beq {\begin{equation}}
\def \eeq {\end{equation}}

\newcommand{\Qstack}{{\hat{\bf Q}}_{\rm stack}}
\newcommand{\hatcone}{{\hat{\bf c}}}
\newcommand{\Ahat}{{\hat{\bf A}}}
\newcommand{\Bhat}{{\hat{\bf B}}}
\newcommand{\Chat}{{\hat{\bf C}}}
\newcommand{\half}{\frac{\footnotesize{1}}{\footnotesize{2}}}
\def \ss {{\bf s}}
\def \SS {{\bf S}}
\def \kkLT {{\bf Q_{\rm LT}}}
\def \la {{\langle}}
\def \ra {{\rangle}}

\newcommand{\tsvec} {{\tilde{\svec}}}
\newcommand{\svec} {{\bf s}}
\newcommand{\kvec} {{\bf k}}
\newcommand{\Qvec} {{\bf Q}}
\newcommand{\rr} {\rvec}
\newcommand{\RR} {\rvec}
\newcommand{\nn} {{\bf n}}
\newcommand{\kk} {{\bf k}}
\newcommand{\LL} {{\bf L}}
\newcommand{\qq} {{\bf q}}
\newcommand{\QQ} {{\bf Q}}
\newcommand{\MM} {{\bf M}}
\newcommand{\rvec} {{\bf r}}
\newcommand{\HH}   {{\cal H}}
\newcommand{\eqr}[1]{(\ref{#1})} 
\newcommand{\tJ}{{\tilde{J}}}

\title{Nonplanar ground states of 
        frustrated antiferromagnets on an octahedral lattice}

\author{Sophia R. Sklan}
\affiliation{Laboratory of Atomic and Solid State Physics, Cornell University,
Ithaca, New York, 14853-2501}
\affiliation{Department of Physics, Massachusetts Institute of Technology, 
Cambridge, Massachusetts 02139}
\author{Christopher L. Henley}
\affiliation{Laboratory of Atomic and Solid State Physics, Cornell University,
Ithaca, New York, 14853-2501}


\begin{abstract}
We consider methods to identify the classical ground state for
an exchange-coupled Heisenberg antiferromagnet on a
non-Bravais lattice with interactions $J_{ij}$ to several neighbor distances.
Here we apply this to the unusual ``octahedral'' lattice 
in which spins sit on the edge midpoints of a simple cubic lattice.
Our approach is informed by the eigenvectors of $J_{ij}$
with largest eigenvalues. 
We discovered two families of non-coplanar states:
(i) two kinds of commensurate state with cubic symmetry,
each having twelve sublattices with spins pointing in (1,1,0) directions
in spin space (modulo a global rotation);
(ii) varieties of incommensurate conic spiral.
The latter family is addressed by projecting the three-dimensional
lattice to a one-dimensional chain, with a basis of two (or more) sites
per unit cell.
\end{abstract}


\maketitle

\section{Introduction}

This paper concerns the classical ground state of the Hamiltonian
  \begin{equation}
  \HH  =\underset{ij}{\sum}
     - J_{ij}\svec_{i}\cdot\svec_{j}
  \label{eq:Ham}
  \end{equation}
where $\{ \svec_i \}$ are unit vectors, and the couplings 
$\{J_{ij}\}$ have the symmetry of the lattice and may extend 
several neighbors away (being frustrated in the interesting cases).

After an antiferromagnet's ordering pattern (or partial information)
is determined by neutron diffraction, the next question is which
spin Hamiltonian(s) imply that order, if we admit interactions $J_2$ 
to second neighbors or $J_n$ to further neighbors.
The starting point for understanding ordered states
is always the classical ground state(s).
If the spins sit on a {\it Bravais} lattice (e.g. face-centered cubic), 
the solution is trivial due to a rigorous recipe, called the 
``Luttinger-Tisza" (LT) method (see Sec.~\ref{sec:LT} below):
the spins adopt (at most) a simple spiral -- a {\it coplanar}
state, meaning all spins point in the same plane 
of spin space~\cite{spiral59,Kapl06}.
But if the spins form a lattice with a basis 
(more than one site per primitive cell),
-- e.g. kagom\'e, diamond, pyrochlore, or half-garnet lattices -- 
no mechanical recipe is known to discover the ground state.
In these more complicated lattices, 
magnetic frustration (competing interactions)
often induces complicated spin arrangements.

Our aim has been to find a recipe 
for general lattices (albeit neither exhaustive nor rigorous)
to discover the ground state spin pattern corresponding to a 
given set of exchange couplings $J_i$, to neighbors
at successive distances.
That is obviously a prerequisite for solving the inverse problem 
(given the ordering patterns found by neutron diffraction,
which combination(s) of interactions can explain them?).
Furthermore, after the whole phase diagram is mapped out,
we can identify the parameter sets leading to exceptionally
degenerate or otherwise interesting states, so as to
recognize which real or model systems might be close to
realizing those special states.

\SAVE{
If the number of degenerate degrees of freedom is extensive
we have a ``highly frustrated'' state.
Such systems are likely to realize novel spin-liquid states
the quantum problem with spin length of 1/2 or 1. 
In  a classical system, the phase diagram (including
anisotropies) is likely to be very rich near to such a 
special point, since small perturbations can select 
many different ordering patterns.}

In this work, we focus on a narrower question:
which parameter combinations
give a {\it noncoplanar} ground state,
which could never happen in a Bravais lattice?
We adopt the exchange Hamiltonian \eqr{eq:Ham},
excluding single-site anisotropies and
Dzyaloshinskii-Moriya couplings,
which can trivially give non-coplanar ground states.

We do not count cases where a non-coplanar ground state
belongs to a degenerate family of states that
also includes coplanar ground states.
This happens trivially when two sublattices
aren't coupled at all, or nontrivially
when the interactions are constrained to cancel.
In the latter cases
thermal or quantum fluctuations usually break the 
degeneracy, favoring the collinear or coplanar states
\cite{shen82,henley89}.
(A small amount of site-dilution or bond disorder
can generate a uniform effective Hamiltonian that 
favors non-coplanar states~\cite{henley89,larson}, 
but here we only consider undisordered systems.)

\SAVE{We also limit the study to highly -- say cubic -- symmetries, 
and ask that all spin sites be symmetry equivalent; again,
it is relatively easy to get non-coplanar states when there
is less symmetry.}

\subsubsection*{Motivations for noncoplanarity}

There are specific physical motivations to hunt for non-coplanar states.
First, they point to possible realizations of {\it chiral}~\cite{Wen89}
spin liquids, such as are described within bosonic large-$N$ formalisms
(as are hoped to approximate the behavior of frustrated
magnets with $s=1/2$). 
\OMIT{Chiral spin liquids were not found in the phase diagram of the
$J_1$-$J_2$-$J_3$ model on the square lattice~\cite{sachdev}.}
Such formalisms describe transitions from an ordered state
to a quantum-disordered spin liquid;
\OMIT{driven e.g. by reducing the
parameter analogous to the spin length.}
since there is no generic reason 
for a state to stop being chiral at the same time it loses spin order, 
a chiral ordered state presumably transitions into a chiral spin liquid.
Hence, as a rule of thumb, a chiral spin liquid is feasible 
if and only if the classical ordered state (on the same lattice)
is non-coplanar.~\cite{sachdev-pc,messio-chiral-liquid}.

Secondly, spin non-coplanarity in metals
(usually induced by an external magnetic field)
allows the {\it anomalous Hall effect}
observed in pyrochlore and other magnets.~\cite{Ta01,anomHall,taillef06,kalitsov08}
This is ascribed to spin-orbit coupling and the 
Berry phases of hopping electrons
(which are zero in the collinear or coplanar case).
\SAVE{In the known materials, the chirality is due to
anisotropies, in contrast to our models where it
is due to exchange interactions alone.}

Thirdly, the symmetry-breaking of noncoplanar 
exchange-coupled magnetic states is labeled by
an order parameter which is an $O(3)$ matrix, 
so the order-parameter manifold is disconnected.
This permits a novel topological defect: 
the $Z_2$ domain wall~\cite{henley84a}, which 
is only possible in non-coplanar phases.
\SAVE{Another topological defect, the $Z_2$ vortex~\cite{Do07}
is possible in any noncollinear phase.)
(cite H. Kawamura, on $Z_2$ vortices in triangular AFM).}

Finally, there is current interest in ``multiferroic'' materials
(i.e. those with cross couplings of electric and magnetic polarizations).
For example, in the canonical multiferroics RMnO$_3$ (where R=rare earth),
frustrated exchange interactions induce a coplanar spiral, which
in the presence of Dzyaloshinskii-Moriya anisotropic interactions 
carries an electric polarization 
with it~\cite{kimura06,kimura-review,cheong-mostovoy,khomski09,kaplan-CoCr2O4}.
If these spirals were asymmetric conic spirals, like our second
class of ground states, 
there is generically a net moment along the axis,
which serves as a convenient ``handle'' to externally manipulate 
the orientation of the ground state (and thus control the multiferroic properties).

\subsection{The octahedral lattice}
\label{sec:phase-octahedral}

Our spins sit on a rarely studied lattice we christen  the
``octahedral lattice'', consisting of the 
medial lattice (bond midpoints) of a simple cubic lattice.
thus forming corner-sharing octahedra.  
Thus, each unit cell has a basis of three sites, forming what we call the
$x$, $y$, and $z$ sublattices (according to the direction of the bond
they sit on). Each cubic vertex is surrounded by an octahedron of 
six sites, with nearest-neighbor bonds forming its edges;
these octahedra share corners, much as triangles or
tetrahedra share corners in the well-known kagom\'e and pyrochlore lattices.
(Indeed, although the ``checkerboard'' lattice was introduced as a two-dimensional
version of the pyrochlore lattice~\cite{moessner98}, 
the octahedral lattice is the best three-dimensional generalization
of the checkerboard lattice.)
This lattice was first studied as a frustrated
Ising antiferromagnet~\cite{chui77,reed77}
More recently, it was used as a (simpler/pedagogical) toy model 
in papers aimed at the ``Coulomb phase'' 
of highly constrained spins 
on a pyrochlore lattice~\cite{hermele04,pickles08}.
It is one of the lattices constructed from the root lattices of Lie algebras.~\cite{shankar}

The octahedral sites are Wyckoff positions (and hence candidates for a 
magnetic lattice) in most cubic space groups, so this is plausible to find in
real materials, and a few are known.
Most simply, it is realized by the transition
metal sites in the Cu$_3$Au superstructure of the
fcc lattice~\cite{chui77} 
(i.e. all but one of the four simple cubic sublattices).
So far, the only example known in which the 
``Cu'' lattice is magnetic seems to be Mn$_3$Ge
which is ferromagnetic~\cite{takizawa02}
Another  realization is in metallic perovskites
such as Mn$_3$SnN, though again the known
materials are ferromagnetic~\cite{fruchart}.
It would actually seem quite plausible to find
realizations of our models, which mostly have several
exchange interactions $J_i$ with competing signs,
among metallic alloys:
the RKKY interaction, expected between local moments in any metal,
oscillates with distance inside a slowly decaying envelope.

The octahedral lattice is closely related to the magnetic lattice 
found in the (mostly metallic) Ir$_3$Ge$_7$ structures, including 
the strong-electron-interaction superconductor Mo$_3$Sb$_7$~\cite{Ko08,Tran08}.  
In that lattice, the simple-cubic
lattice sites are surrounded by {\it disjoint} octahedra,
i.e. a dimer of {\it two} magnetic ions decorates each bond of 
the simple cubic lattice.
If this dimer were strongly coupled ferromagnetically, it would
be a good approximation to treat it as a single spin,
which is exactly the octahedral lattice.
Instead, in Mo$_3$Sb$_7$ the 
dimers are antiferromagnetically coupled 
and, since Mo has spin 1/2, they form singlets~\cite{Tran08}.
If the spin length were longer, justifying classical treatment, 
we could convert to the ferromagnetic case simply by inverting the spin
directions in every octahedron around an odd site of the
cubic lattice, and changing the sign of all bonds coupling
even sites with odd sites.  Thus, much of the classical phase diagram 
for the Mo$_3$Sb$_7$ lattice is related to that of the octahedral
lattice.

In this paper, we mainly consider four kinds of couplings,
for separations out to the third neighbors:
$J_1$ for $\langle 1/2, 1/2, 0 \rangle$,
$J_2$ or $J_2'$ for $\langle 1,0,0 \rangle$,
$J_3$ for $\langle 1, 1/2, 1/2 \rangle$.
Notice that 
couplings with the same displacement
need not be equivalent by symmetry, since the site
symmetry is just fourfold, less than cubic.
(Our naming convention is to use the prime for the separation
which requires more first-neighbor steps to traverse.)
We also (less extensively) consider 
interactions $J_4$ or $J_4'$ for $\langle 1,1,0\rangle$.
To organize our exploration of this parameter space, 
in analytic calculations we shall often assume 
$J_3, J_4, J_4' \ll J_1, J_2, J_2'$
(that suffices to give examples of most of the 
classes we found of noncoplanar ground states).

\subsection{Outline of paper and preview of results}

We begin (Sec.~\ref{sec:framework})
by developing the techniques and concepts
necessary to find the phase diagram as a function of the $J_{ij}$'s
and to discover {\it non-coplanar} ground states.
 We found ground states using three methods.
The first (Sec.~\ref{sec:LT}
was Fourier analysis, known as the ``Luttinger-Tisza'' method,
which can give a lower bound on the energy, but may not
give a full picture of the ground state. 
The second (Sec.~\ref{sec:iter-min}
is an iterative minimization algorithm, 
which numerically converges to a ground state; 
we introduce several diagnostic tools for
understanding the spin patterns produced by iterative minimization.
The third method (Sec.~\ref{sec:var}) is the variational optimization of 
idealized patterns displayed by iterative minimization.

We then turn to our results, beginning with descriptions of the 
several classes of magnetic state we found for the octahedral lattice:
various coplanar states (Sec. ~\ref{sec:coplane}), 
the noncoplanar, commensurate ``cuboctabedral'' spin states 
(Sec.~\ref{sec:cuboc}), and 
a more generic group of noncoplanar, 
incommensurate ``conic spirals''
(Sec.~\ref{sec:conic}); 
in these the lattice breaks up into layers of
spins with the same directions, each layer 
being rotated around the same (spin-space) 
axis relative to the layer below. 
Particularly noteworthy was a ``double-twist'' state
we encountered, which is something like a conic
spiral which also has a complex modulation 
in the transverse directions (Sec.~\ref{sec:doubletwist}).
The plain stacked structures can be studied by 
mapping to one-dimensional
(``chain'') lattices, also with couplings to many neighbors,
as worked through in Sec.~\ref{sec:mapping}.

From this we go on (Sec.~\ref{sec:phase})
to quickly survey the phase diagrams we found,
first for the cuboctahedral lattice, and then
for the chain lattice (when treated as a lattice
in its own right).
In the conclusion, Sec.~\ref{sec:discussion},
we reflect on what our results might suggest 
for other frustrated lattices, such as the pyrochlore.

\section{Methods and framework}
\label{sec:framework}

\SAVE{In this paper, we employ 3 techniques -- note carefully the
terminology: (1) iterative minimization, (2)
numerical optimization of variational form (which may be done by 
sampling a grid of parameter values) and (3) analytic optimization
of the variational form.}

We employed several approaches to discover and
understand ground states, for each given set of
interactions (these are developed rest of this section --
except for the use of mappings, which we explain in Sec.~\ref{sec:conic}, where it becomes natural to employ this technique).

\begin{itemize}
\item[(a)]. A Fourier analysis of the Hamiltonian \eqr{eq:Ham}
as a quadratic form with coefficients $J_{ij}$, the so-called
``Luttinger-Tisza'' method outlined in Sec.~\ref{sec:LT}.

\item[(b)]. Iterative minimization, our main ``exploratory'' technique.
Starting from a random initial condition, 
we successively adjusted randomly
chosen  spins so as to reduce the energy.
(Sec.~\ref{sec:iter-min}).
We then analyzed each resulting pattern 
with various diagnostics, as described in
Sec.~\ref{sec:iter-min}, and tagged
 the non-coplanar ones for further investigation.

\item[(c)]. Variational optimization of the 
iterative minimization ground state. Finding a closed-form for
the ground state introduces a number of free parameters (the most
obvious being a wave-vector). By allowing these parameters to vary from
the values found with iterative minimization, we find a new, more rigorous, ground state.

\item[(d)]. Mapping the (three-dimensional) problem 
to a similar problem in a one-dimensional ``chain'' lattice 
with a basis of two sites.  This is valid when the
optimal (three dimensional) spin configuration 
is a stacking of layers, which we judged based on the results 
from approaches (a) and (b).
The states on this simplified chain lattice 
may be found using approaches (a) and (b), 
or analytically solved after parametrizing the state
with a set of variational parameters.

\end{itemize}

\subsection{Spin states and Luttinger-Tisza modes}
\label{sec:LT}

The general theory of spin arrangements is reviewed in
Refs.~\onlinecite{Kapl06} and \onlinecite{naga67}.
The most fruitful approach to finding the ground states of the 
Hamiltonian \eqr{eq:Ham} is to treat it as a quadratic form
rewriting \eqr{eq:Ham} as
   \begin{equation}
       \HH = - \sum_{\kvec}\sum_{\alpha, \beta=1}^m
       \tJ_{\alpha\beta}(\kvec) \tsvec_\alpha(-\kvec)\cdot\tsvec_\beta(\kvec),
   \label{eq:Ham-fourier}
   \end{equation}
where $\alpha$ and $\beta$ are sublattice indices; the explicit
formulas for the cuboctahedral lattice case are given in
Appendix \ref{sec:pert-LT}, Eq.~\eqr{eq:LT-matrix}.
Then we diagonalize this matrix, obtaining
   \begin{equation}
       \HH = - \sum_{\kvec}\sum_{\nu=1}^m \tJ(\kvec \nu) |\tsvec(\kvec\nu)|^2.
   \label{eq:modes-fourier}
   \end{equation}
Here $m=3$ is the number of sites per primitive cell, and
$\nu$ is a band index; thus
$\{ \tJ(\kvec\nu) \}$ are the eigenvalues of $J_{ij}$
as an $mN\times mN$ matrix, with $N$ being the number of cells
(to be taken to infinity), and
the wavevector $\kvec$ runs over the Brillouin zone.
[In the Bravais lattice case, $m=1$, 
the eigenvalue $\tJ(\kk)$ is simply the Fourier transform of $J(\RR)$;
for $m>1$.]
Also, $\tsvec(\kvec\nu)$ (complex-valued 3-vector)
is the projection of the spin configuration
onto the corresponding normalized eigenmode,
$N^{-1/2} v_{\nu s} \exp i (\kvec\cdot\rvec)$.
We shall call these the ``Luttinger-Tisza'' (LT) eigenvalues
and modes~\cite{Kapl06,LT,Ly60};
a mode with the most negative $\tilde{J}(\kk\nu)$ 
is called an ``optimal'' mode, and
its wavevector is called $\{ \kkLT \}$.

The ideal case is that we can build a spin state satisfying two conditions
\begin{itemize}
\item[]{\it Condition (1)}
$\{\ss_i\}$ are entirely linear combinations of optimal LT modes 
\item[]{\it Condition (2)}
$|\ss_i|^2=1$ everywhere (unit length constraint)
\end{itemize}
If both conditions are satisfied, these {\it must} be ground states, 
and all ground states {\it must} be of this form.

In the case of a Bravais lattice ($m=1$),
the LT modes are just plane waves $e^{i\kk\cdot \rr}$, 
and one can always construct a 
{\it planar} spiral configuration~\cite{spiral59},
$\svec(\rr) = \cos (\kkLT\cdot \rr) \Bhat +  \sin (\kkLT\cdot \rr) \Chat$, 
where $\Bhat$ and $\Chat$ are orthogonal unit vectors, 
and the spatial dependence consists only of optimal modes~\cite{Kapl06}.
In the simplest cases, $\kkLT$ is at high
symmetry points on Brillouin zone corners,
and one can construct a combination of optimal 
modes which is $\pm 1$ on all sites, 
which defines a {\it collinear} ground state, 
as in the phase diagrams in Ref.~\onlinecite{Sm66}.

Thus, non-Bravais lattices are {\it necessary} in order
to get non-coplanar states.  (But not sufficient:
it appears that, on non-Bravais lattices with high
symmetry, the commonest ground states are still
collinear or coplanar.)
In lattices-with-a-basis, however, the LT eigenmodes
have different amplitudes on different sites within
the unit cell, and it is not generally possible
to make any three-component linear combination of
the best modes that satisfies the unit-length constraint. 
(There is an exception for lattices in which 
the neighbors-of-neighbors are all second neighbors,
such as the diamond~\cite{Be07b} or honeycomb~\cite{honeycomb}
lattices.

A ``generalized'' L-T method for non-Bravais lattices
was introduced by Ref.~\onlinecite{Ly62} (see also Ref.~\onlinecite{Fr61})
and applied to spinels with both A and B sites magnetic \cite{Ly60,Kapl06}.
However, this method involves site-dependent variational parameters,
so one must already understand the pattern of the ground state in 
order to make it into a finite problem; in practice, this method
appears quite similar to our method 
(Sec.~\ref{sec:conic} and \ref{sec:pert-state}) 
of projecting a layered structure to a one-dimensional chain.


\SAVE{In certain special cases
the LT method works on non-Bravais lattices with $J_1$ and $J_2$,
e.g. the diamond lattice~\cite{Be07b} representing an A-spinel.
Namely, if the first-neighbors-of-first-neighbors are
all second neighbors (except one which is the original site).
In  such cases, the set of optimal wavevectors $\{\kk_{LT} \}$
a distorted circle/sphere in reciprocal space \cite{Be07b}.}

Although the LT optimal modes (usually) give the 
{\it exact} ground states in the cases we focus on, we believe 
the exact ground state is frequently built mainly from almost-optimal
modes; that is, although a linear combination of optimum LT modes 
violates the unit-spin constraint, with a small distortion 
it may satisfy the constraint and be the ground state.
(That distortion necessitates admixing other modes but
with small amplitudes, since they carry a large energy penalty,
according to \eqr{eq:modes-fourier}.)
In particular, we anticipate that (for incommensurate
orderings) the true ordering wavevector lies in 
the same symmetry direction as the LT wavevector;
and that the phase diagram for optimum LT wavevectors 
mostly has the same topology as the actual phase diagram 
for ground states.
Thus the LT modes can serve as  a ``map'' 
for navigating the parameter space of $\{ J_i \}$ 
and for understanding the ground state spin configurations.

An important caveat is that almost all $\QQ$ vectors
have symmetry-related degeneracies, and the LT analysis
is silent on how these modes are to be combined with
different spin directions, so the specification of the
actual spin configuration is incomplete. 
(An example is the ``double-twist'' state, of Sec.~\ref{sec:doubletwist}.)
As a corollary, a single phase domain on the LT mode phase diagram 
might be subdivided into several phases in the spin-configuration phase
diagram, that represent different ways of taking linear combinations
of the same LT modes. This cannot be detected at the LT level.

One immediate insight is afforded by considering the LT phase
diagram.  Short-range couplings have Fourier
transforms $\tJ_{\alpha\beta}(\kk)$  in \eqr{eq:Ham-fourier}
that vary slowly in reciprocal space. Such functions 
typically possess extrema at high-symmetry points in the Brillouin zone;
the same is probably true for the optimum {\it eigenvalues}
and their wavevectors $\kkLT$.  That corresponds to simple,
commensurate ordering in real space.  In order to get the 
optimal LT mode (and presumably the actual ordering)
to be incommensurate, or to possibly stabilize 
states with stacking directions other than (100), 
one needs to include more distant neighbor couplings.

In practice, we never used LT to directly discover
the ground state spin configuration; its value is to quickly
prove a given state is a ground state.
But the LT viewpoint did inform the Fourier-transform diagnostic we 
used in analyzing the outputs of iterative relaxation (Sec.~\ref{sec:iter-min}.
Furthermore, when we operated in the ``designer'' mode 
(seeking the couplings that stabilize a specified state)
we used the LT modes as a guide or clue:
namely, we found the $\{ J_i \}$ that
made the ordering wavevector $\QQ$ of our target state
to be the optimal $\kkLT$, which is easier than making 
$\QQ$ be the ordering wavevector of the
actual ground state.

\subsection{Iterative minimization}
\label{sec:iter-min}

Our prime tool for exploration was iterative minimization 
starting from a random initial condition.
Random spins are selected in turn
and adjusted (one at a time) so as to minimize the energy,
by aligning with the local field of their neighbors, 
till the configuration converged on a local minimum
of the Hamiltonian.~\cite{WW}.
(Our criterion was that the
energy change in one sweep over the lattice was less than 
a chosen tolerance, typically $10^{-9}$).

It might be worried that such an algorithm gets stuck in
metastable states, unrepresentative of the ground state; such
``glassy'' behavior is indeed expected in the case of Ising 
(or otherwise discrete) spins, or in {\it randomly} frustrated
systems such as spin glasses.  However, vector spins
typically have sufficient freedom to get close to the true
ground state~\cite{henley84b,henley-hfm00}.  
The typical ways they deviated from the ground state
are just long-wavelength wandering (``spin waves'') or 
twists of the spin directions.

The only problem with the dynamics is that our algorithm 
is a variant of ``steepest descent'', one of the slowest of 
relaxation algorithms.   
such deviation modes are indeed slow relaxing 
For this kind of (local) dynamics, the relaxation rate 
of a long-wavelength spin wave at wavevector $\qq$
is proportional to $|\qq|^2$, i.e. $1/L^2$ for the
slowest mode in a system of side $L$.
(In future applications, some version of conjugate gradient
should be applied to give a faster convergence, or -- 
if there is a  problem is finding the right valley of 
the energy function --  one might adapt Elser's ``difference map'' 
approach to global optimization~\cite{El07}.)

For initial explorations, we usually
used very small cubic simulation boxes of $L^3$ cells
($L=3$, 4, or 5).
For each set of $\{ J_{ij} \}$ tested, we tried 
both periodic and antiperiodic boundary conditions, as well
as even or odd $L$. Usually one of those four cases
accomodates an approximation of the infinite system 
ground state, though of course any incommensurate
state must adjust either by twisting to shift the
ordering wavevector to the nearest allowed value, 
or else (as we observed) via the formation of wall defects. 
We tried to distinguish the ground states which were
``genuine'' in that a similar state would remain stable 
in the thermodynamic limit.  In particular, out of the
four standard systems we tried (even/odd system size, 
periodic/antiperiodic BC's), a ``genuine'' state should
be the one with lowest energy.

For a large portion of the parameter space, the ground
states were planar spirals, essentially no different 
from the solutions 
guaranteed in the Bravais lattice case.
Many of the noncoplanar configurations found were
``non-genuine''
artifacts of finite size when the
periodic (or antiperiodic) boundary  conditions 
and dimension $L$ were incompatible with the 
natural periodicity of the true ground state.  
One might expect the wrong boundary
condition to simply impose a twist by $\pi/L$
per layer on the true ground state, but instead
the observed distortion of the spin texture
was sometimes
of the natural periodicity of the ground state,
a ``buckling'' occurred; that is, the configuration consists of finite domains
similar to the true ground state, separated by soliton-like 
domain walls.  

The greatest difficulty in our procedure was not obtaining an 
approximation of the ground state, or even deciding whether it was genuine.
Rather, it was grasping what the obtained pattern is, 
and how to idealize it to a periodic (or quasiperiodic)
true ground state of the infinite system.  
We were aided by the following three diagnostics.

\subsubsection{Diagnostic: Fourier transform:}

Configurations obtained by iterative minimization were Fourier transformed
and the norms of each Fourier component $\sum_s |\tsvec(\kk,s)|^2$
were summed (combining the sublattices)~\cite{FN-strucfactor}.
This suffices to identify the state when it is a relatively
simple antiferromagnetic pattern, or an incommensurate 
state described as a layer stacking.
In any case, the results can be compared to the LT
mode calculation to see if the found state achieves the
LT bound.

\subsubsection{Diagnostic: common-origin plot:}

\begin{figure}
\includegraphics[scale=0.4]{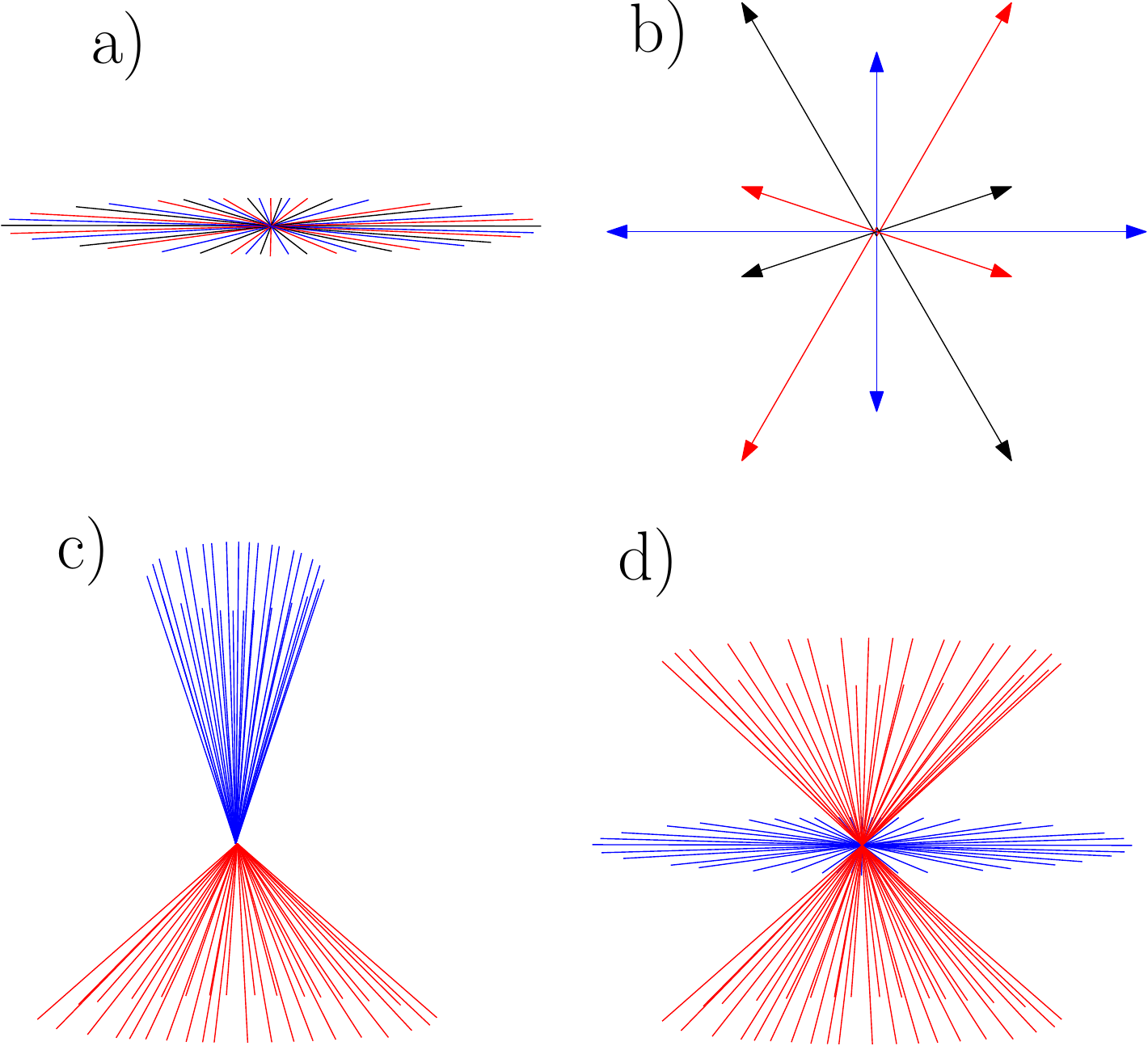}
\caption
{(COLOR ONLINE). Common-origin plots for four kinds of
spin ground states on the octahedral lattice,
appearing for different choices of the exchange
interactions $\{ J_i \}$.
(a). planar spiral
(b). cuboctahedral state
(c). asymmetric conic spiral
(d). alternating conic spiral.
}
\label{fig:cop}
\end{figure}

The simplest visual diagnostic of a state
is the "common-origin plot", in which each
spin's orientation is represented as a
point on the unit sphere.  For example,
an incommensurate
coplanar spiral state would appear as a single
great circle on the common-origin plot 
(see Fig.~\ref{fig:cop}).

A drawback of the common-origin plot is the lack of
information on the spatial relation of the spins.
(For example, a ``cone'' might appear consisting of
closely spaced spin directions, but these might belong
to widely spaced sites.) Furthermore, this diagnostic
is quite fragile in configurations where a domain wall
or other defect has been quenched in.

In the case of one-dimensional chains (see Sec.~\ref{sec:mapping}), 
we may instead use the ``end-to-end'' spin plot, 
where the tail of each spin vector is on the head
of the previous spin vector.  The advantage is that
(i) images are not so obscured by overlaying of
different vectors, and (ii) spatial information is
captured, in particular defects where the state has
``buckled''.~\cite{FN-higherD}

In tandem with the common-origin plot, an algorithm was
used that found groups of spins 
with (nearly) identical spin directions, 
within a chosen tolerance (typically, a dot product greater than 0.99).  
For commensurate patterns comprising a finite set of spin directions,
this allowed the magnetic unit cell and spin configuration
to be read off, but it was useless for incommensurate
states, in which no spin direction exactly repeats.

\subsubsection{Diagnostic: Spin moment-of-inertia tensor}

We computed the $3\times 3$ tensor 
  \begin{equation}
     M_{ij} \equiv \frac{1}{mN} 
     \sum _{\alpha,r} (\svec_i(\vec{r}))_\alpha (\svec_j(\vec{r}))_\alpha
  \end{equation}
(notice ${\rm Tr}(\MM)=1$) and diagonalized it.
We recognize coplanar or non-coplanar spin states as
those where $\MM$ has two or three nonzero eigenvalues, respectively.  
(Rotating the spin configuration so the principal
directions of $\MM$ are the coordinate axes usually
manifests the spin configuration's symmetries.)

\subsection{Variational Optimization}
\label{sec:var}

Through the diagnostic techniques described in the previous subsections,
it is normally straightforward to parameterize the spins in the ground as $\svec(\RR,\{\alpha\})$, i.e. as a function of position and some arbitrary set of parameters, $\{\alpha\}$. The exact values of these parameters can (formally) be calculated from optimizing $\HH(\svec(\RR,\{\alpha\}),\{J\})$. One major advantage of relying upon this method
is that it reduces the effect of numerical artifacts from iterative minimization. For example, we no longer enforce an arbitrary periodicity upon the LT wavevector.

\subsection{Conceptual framework for bridging states}
\label{sec:bridging}

This subsection is not about a technique, but a classification of
two ways that ground states may be related to each other,
and thus of two kinds of phase boundaries in the phase diagrams 
(Sec.~\ref{sec:phase} and Appendix~\ref{app:phase-boundaries}).
We call these two concepts
``encompassing states'' and ``families of degenerate states''.

\subsubsection{Encompassed states}
\label{sec:encompassed}

We call a ground state ``encompassed'' if it is a special case of 
another, more  general state. For example, a ferromagnet is 
encompassed by a helimagnet, since letting the helimagnetic angle 
go to zero produces a ferromagnet.   
``More general'' means there is a continuous family of states 
such that each particular
combination of couplings $J_i$'s completely determines a 
particular member of that family.
Moreover, the (more general) encompassing state necessarily spans at least one more 
dimension of spin space, so if the encompassed state is coplanar, 
the encompassing one is non-coplanar.
However, while every encompassing state is more general than the state it encompasses, not every more general state will encompass a particular ground state. For example, the asymmetric conic is more general than the splayed ferromagnet 
(both of these states are defined in Sec.~\ref{sec:conic}), 
but it does not encompass the splayed ferromagnet; there is, 
however, a third class that encompasses both these classes.

\subsubsection{Degenerate states}

{\it By contrast, ``degenerate states'' means that for certain 
combinations of $J_i$'s, there is a continuous family of 
exactly degenerate ground states.}
Most commonly, this is the result of {\it decoupling} between 
sublattices of spins, meaning one can apply a global rotation limited to 
just one of the sublattices while remaining in the degenerate manifold.  
This can come about in two ways.
The trivial way is when all $J_i$'s that couple those sublattices
vanish.
The more interesting way is when the couplings are nonzero,
but cancel generically in all the ground states; the simplest example 
of this kind is the $J_1$--$J_2$ antiferromagnet on the square~\cite{henley89} 
or bcc~\cite{shen82} lattice, in the $J_2$-dominated regime 
in which each of the even and odd sublattices realizes plain
N\'eel order.  Apart from decoupling, degenerate manifolds
are also sometimes realized by simultaneous rotations involving
all sublattices with some mutual constraint (e.g. in the nearest-neighbor
kagom\'e lattice, the constraint that the spins add to zero
in every triangle).

Independent of the categories mentioned, these degenerate families may
either be {\it simply degenerate}, meaning the ground state manifold is labeled
by a finite number of parameters -- one nontrivial angle in the case of 
the $J_1$--$J_2$ antiferromagnet -- 
-- or else {\it highly degenerate}, 
meaning the number of parameters scales with size as $O(L^n)$ with $n>0$.
An example with trivially decoupled, highly  degenerate ground states
is a layered lattice with vanishing interlayer couplings, so the number of free
parameters scales as $O(L)$; the $J=1$ kagom\'e antiferromagnet
is a mutally constrained, highly degenerate case with $O(L^2)$ parameters, 
i.e. extensively many.

Both the simply and highly degenerate families also have
clear signatures in reciprocal space. 
In the simple case, the optimal wavevectors are discrete
and related by symmetry, 
e.g. the $J_1$--$J_2$ square lattice antiferromagnet has 
$\kkLT= (1/2,0)$  or $(0,1/2)$;
in this case, the degeneracy lies in the freedom to mix these degenerate
LT eigenmodes with different coefficients,
not from the presence of extended (one-,  two-, or even three-dimensional) surfaces in the Brillouin zone.
By contrast, in the highly degenerate scenario, 
typically holding for special combinations of couplings, 
the LT optimum wavefectors occur not just at isolated $\kkLT$ in the
Brillouin zone, but on extended (one-,  two-, or even three-dimensional) 
surfaces, i.e. one has degenerate eigenmodes of the LT matrix that are
not symmetry-equivalent.
The rhombohedral lattice with $J_1$, $J_2$, and $J_3$ has a 
degenerate one-parameter family of wavevectors corresponding to
different coplanar spirals~\cite{rastelli-tassi}.
The three-dimensional pyrochlore lattice with only nearest-neighbor 
($J_1)$ couplings is a well-known example of the highly degenerate 
scenario, requiring extensive number of parameters.
In that case, the minimum $LT$ eigenvalue is uniform throughout
the Brillouin zone~\cite{Rei91} (a so-called ``flat band'').



\subsubsection{Encompassed and degenerate states as bridges in phase diagram}

What encompassing states and families of 
degenerate states have in common 
is to serve as bridges between simple states.

In the ``encompassing'' case, the encompassing state is typically
stable in a domain of parameter space of nonzero measure.
When one adjacent phase in a phase diagram is encompassed by the
other, they are necessarily related (in our $T=0$ phase diagram)
by a continuous transition, usually involving a symmetry breaking.

In contrast, degenerate families are (frequently) confined to phase
boundaries. Even when they occupy a finite
area in a slice of parameter  space (e.g. the $(J_1,J_2)$ plane when
all other couplings are zero), turning on additional couplings
can remove the degeneracy.  

A corollary is that the naive classification
of continuous or first-order phase transitions does not work.
Consider two phases separated by a phase boundary on which a degenerate
family is stable.  Each of the two phases (or the limit of either
as the boundary is approached) is a special case from the
degenerate family.  Since the limits taken from the opposite
directions are different, it appears at first as an abrupt 
transition.  On the other hand, it is possible to take the system
continuously from one phase to the other if we pause the parameter
variation when we hit the phase boundary, and follow a path through 
the degenerate manifold from one of the limiting states to the other
one.  

Furthermore, turning on additional parameters generically
destroys the degeneracy.  That converts the degenerate family into an 
encompassing family, and the single phase boundary into
two continuous ones.  Specifically, starting in one of the main phases, we cross 
a small strip of phase diagram in which the configuration evolves 
(determined by the parameter combination) from one of the limiting states to the
other one, and then enter the other of the main phases.
Thus, the ``encompassed'' kind of transition is distinct from either
a first-order transition (between two unrelated states) or an ordinary
continous one, and will be indicated on phase diagrams with a distinct
kind of line.

\subsection{Cluster analysis: two degenerate ground states}
\label{sec:cluster}

The ``cluster'' method is a rigorous analytic approach to 
ground states, alternative to the LT mode approach.~\cite{Ly64},
which depends on decomposing the Hamiltonian into terms for
(usually overlapping) clusters, and finding the ground states for one
cluster.  If these ground states can be patched together so as 
to agree where they overlap, the resulting global state must be 
a ground state and all ground states must be decomposable in
this fashion.
In this way we can characterize the degenerate states
appearing for two special combinations of $J_i$'s.  

\subsubsection{Antiferromagnetic $J_1$ only}
\label{sec:onlyJ1}

In this case, the cluster is a triangle (one face of an octahedron, including 
one site each from the $x$, $y$, and $z$ sublattices).  
The ground state of such a triangle is the usual 120$^\circ$ 
arrangement of spins. If all such triangles are to be satisfied,
then wherever two of them share an edge, the respective unshared 
spins are forced to have identical directions -- in the
present case, spins on opposite corners of the octahedron.
Thus, a line of $x$ sublattice spins in the $x$ direction
(or similarly of the other sublattices in their directions)
is constrained to be the same.

This high degeneracy is not limited to the single point
in parameter space $J_1<0$. If we turn on $J_2$, which 
couples the nearest neighbors aling those lines of spins,
the same configurations remain the ground state until $J_2$
is negative and its magnitude sufficiently large compared to $|J_1|$;
less obviously, the same thing is true for $J_3=J_4$,
varied together.  
\SAVE{Consider the vertices (0,0,0) and (1,0,0) of the cubic lattice.
Couple the sum of the four $y$ and $z$ spins around (0,0,0)
to the four around (1,0,0).  On the one hand, in this state
both must be proportional to the $x$ spins along that line, 
which are the same.  On the other hand, such combinations
account for all $J_3$ and $J_4$ bonds.}

This allows two different kinds of highly degenerate state:

(a) One sublattice (say $x$) has $\ss_i = +\Ahat$ along every line.  
Within the other sublattices, 
each $yz$ plane has an independent rotation about the $\Ahat$
axis. Thus the spin directions are 
    \begin{equation}
	  \ss_i = -\frac{1}{2} \Ahat \pm \frac{\sqrt 3}{2} \Bhat(x),
    \label{eq:J1rot}
    \end{equation}
where $\Bhat$ is a different unit vector in each plane, and we 
take the $+$ or $-$ sign in the $y$ or $z$ sublattice, respectively.

The common-origin plot for this state looks superficially like a
conic spiral, the cone being formed by the $y$ and $z$ spin directions.
In reality, whereas an incommensurate spiral gives a uniform weight
along the spiral in the common-origin plot, this state gives a
random distribution which approaches uniformity only in the limit
of a very large system.

(b). For a second family of (discretely) degenerate states, we choose
   \begin{subequations}
   \begin{eqnarray}
	\ss(n_1+1/2,n_2,n_3) &=& 
	      \frac{1}{\sqrt{2}} [0, f_2(n_2), -f_3(n_3)]; \\
	\ss(n_1,n_2+1/2,n_3) &=& 
	      \frac{1}{\sqrt{2}} [-f_1(n_1), 0, f_3(n_3)]; \\
	\ss(n_1, n_2, n_3+1/2) &=& 
	      \frac{1}{\sqrt{2}} [f_1(n_1), -f_2(n_2),0].
   \label{eq:J1cuboc}
   \end{eqnarray}
   \end{subequations}
where $f_i(n_i)=\pm 1$ are arbitrary.
Notice that \eqr{eq:J1cuboc} uses (a subset of) the cuboctahedral
directions. Typically, in a sufficiently large system, all those
directions are used nearly equally; the common-origin plot 
would show a cuboctahedron.
However, the spins do not have a regular pattern in space
since \eqr{eq:J1cuboc} is random, with a discrete degeneracy
$O(L)$ in a system of $L^3$ cells.
The states \eqr{eq:J1cuboc} represent a degenerate
family of states, as formulated in Sec.~\ref{sec:bridging}:
the optimum LT eigenvalues are found at {\it all}
wavevectors $\QQ$ lying on the (100) axes.

We are {\it not} interested in the high degeneracy for
its own sake; its significance is that various kinds of
ordered states can be selected out of it, by turning on 
additional couplings (even infinitesimally).
Thus, the high-degeneracy parameter combinations
will be corners of phase domains in the phase diagram.  

\subsubsection{Antiferromagnetic $J_1$ and $J_2$}

Let $\LL_\alpha$ be the net spin of the octahedron
centered on $\alpha$.  
   \begin{equation}
	 \LL_\alpha \equiv \sum _{i \in \alpha} \ss_i
   \end{equation}
where $\alpha$ is a cubic lattice vertex and $i\in \alpha$
means site $i$ is on one of the six bonds from $\alpha$,
Consider a Hamiltonian written as
   \begin{equation}
	 \HH = \frac{J}{2} \sum _\alpha |\LL_\alpha|^2 .
   \label{eq:Ham-octzero}
   \end{equation}
On the one hand, expanding the square shows this
is simply the antiferromagnet with $J_1=J_2=J$. 
On the other hand, it is obvious from \eqr{eq:Ham-octzero}
that any configuration with a net (classical) spin of
zero on every octahedron is a ground  state. This is 
another example of a degenerate ground state family
(Sec.~\ref{sec:bridging});
in this case the continuous degeneracy is macroscopic.
This Hamiltonian is constructed in exactly the same way as those
of well-known highly frustrated lattices (kagom\'e, checkerboard, 
half-garnet, pyrochlore) that have similar ground state degeneracies.

\section{Coplanar states}
\label{sec:coplane}

Several different collinear or coplanar ground states can be
stabilized within the octahedral lattice, only one of which requires
couplings beyond $J_2$. We will describe them from the smallest
to the largest magnet unit cells.
The most elementary of these is the 
ferromagnetic state, in which all spins are aligned in the
same direction, and which 
obviously requires predominantly positive couplings.
This state is composed of a (0,0,0) LT mode with equal 
amplitudes on every sublattice (so the normalization
condition is already satisfied).

There is also the ``three-sublattice 120$^{\circ}$
antiferromagnetic state'', whose unit cell is the
primitive cell.
Each of the three sublattices has a uniform direction;
the net inter-sublattice couplings are antiferromagnetic, so 
(as in the ground state of a single antiferromagnetic triangle)
the respective spin directions are 120$^\circ$ apart and coplanar. 
Thus this state, too, is characterized by 
ordering wavevector $\QQ=(0,0,0)$, but not the same LT mode
as the ferromagnetic state.  Instead, this one is from the two degenerate 
modes at $\QQ=(0,0,0)$ that are orthogonal to the uniform mode.
(Any combination of these modes has unequal magnitude on the different 
sublattices, which is why {\it both} modes need to be present in the
spin state, combined with different spin directions.)
This state is a special case of the highly degenerate ground states
found when only $J_1 < 0$ (Sec.~\ref{sec:onlyJ1}).

The next group of coplanar states are the $\QQ\neq 0$ 
antiferromagnetic states, of which there are three kinds,
characterized by having ordering wavevectors of type
(0,0,1/2), (0,1/2,1/2), or (1/2,1/2,1/2). 
Each of these states is antiferromagnetic overall within every sublattice;
the sublattices decouple,  since any inter-sublattice interaction
couples a spin in one sublattice to equal numbers of spins 
pointing in opposite directions in the other sublattice. 
These are LT states; in the (0,0,1/2) and (0,1/2,1/2) cases, 
the LT mode used is nonzero on only one sublattice, and 
a different one of the three symmetry-related wavevectors
is used for each sublattice (the one with the same distinguished direction).
e.g. the $x$ sublattice uses (1/2,0,0) or (0,1/2,1/2) modes.
Notice that in these two cases, the spins repeat ferromagnetically
along some directions (within a sublattice); this is a consequence
of the anisotropy of the intra-sublattice couplings.
Qualitatively, these states are stable when $J_2$ is different from $J_2'$.
All of these states can be realized with collinear spins.

Lastly, the octahedral lattice admits helimagnetic states. 
These states require at least $J_3$ couplings to become stabilized. The helimagnetic states must be composed entirely out of $(q,q,q)$ modes. This is because any other wave-vector would break the symmetry between the sublattices. More precisely, helimagnetic states are generically a function of one variable, $\vec{k}\cdot\vec{r}$, and are therefore equivalent to a one-dimensional system. It will therefore be amenable to stacking vector analysis, developed in  ~\ref{sec:conic}. But using stacking vectors to transform the octahedral lattice to a one-dimensional chain will necessarily produce a non-Bravais lattice unless the stacking vector (111) (or a permutation of sign). And any helimagnetic mode in the one-dimensional non-Bravais lattice will break normalization in the octahedral lattice, since some spin directions would be represented more than others (this will be allowable for conics because they mix multiple modes, but helimagnets are explictly single mode). Therefore, the only allowable stacking vector (and by implication, wave-vector) is (111).

\section{Cuboctahedral states}
\label{sec:cuboc}

The octahedral lattice possesses two kinds of ``cuboctahedral'' state,
stable in different domains of parameter space, 
for which the common-origin plot takes the form of a cuboctahedron,
i.e. twelve spin directions of the form $(1,1,0)/\sqrt 2$ and its
permutations [Fig.~\ref{fig:cop}(b)].
The magnetic unit cell is $2 \times 2 \times 2$ 
for both of these true cuboctahedral states (spuriously cuboctahedral
states were remarked in Sec.~\ref{sec:onlyJ1}).
They differ in that the angles between neighboring spins 
(which are in different sublattices) is 60$^\circ$ in one kind of
cuboctahedral state but is 120$^\circ$ in the other kind.

As worked through in this section, the cuboctahedral states can be understood 
from any of three approaches:
\begin{itemize}
\item[(a)]
Cluster construction:
the Hamiltonian can be decomposed into a sum of terms, 
each for an octahedron; we can patch together the ground
states of the respective octahedra to obtain a ground
state of the whole lattice.  (For 60$^\circ$ cuboctahedral only.)
\item [(b)] 
Degenerate perturbation theory: 
two special sets of couplings give degenerate 
families of ground states, out of which a small
additional coupling can select the cuboctahedral state.
(For 120$^\circ$ cuboctahedral only.)
\item[(c)]
The Luttinger-Tisza framework of Sec.~\ref{sec:LT}.
(For both kinds of cuboctahedral state.)
\end{itemize}

The 120$^\circ$ cuboctahedral state is a subset of 
the $J_1$--only antiferromagnetic ($J_1<0$) ground states
described above (Sec.~\ref{sec:onlyJ1}).
Thus, this state is stabilized even in the limit as $J_2$ 
(or other distant couplings) become arbitrarily small.
It is the only non-coplanar state we found that does not require
any couplings beyond $J_2$ and $J_2'$.

\subsection{Lattice as union of cuboctahedral cage clusters}
\label{sec:cuboc-cages}

The first cuboctahedral state noticed 
was in the $J_1$--$J_2$ magnet on the kagome
lattice~\cite{Do05,Do07}, with 
$J_1$ ferromagnetic and $J_2$ antiferromagnetic.
There is a range of ratios $J_2/J_1$ in which 
the magnetic unit cell on the kagome lattice is $2\times 2$.  
Taking that cell as given, the possible ground states are
those of the twelve-site cluster made by
giving periodic boundary conditions to one unit cell --
a cluster which is topologically equivalent to a single
cuboctahedron (even when couplings to any distance
are taken into account), and hence include the
cuboctahedral state.
\SAVE{There are actual spirals too, 
running around each hexagon loop of length 6.
(Each hexagon uses a different planar subspace of the spin space).}

Turning to the present case of the octahedral lattice,
in fact this is a union of cuboctahedral cages surrounding 
cube centers, complementary to the octahedral clusters surrounding the cube vertices.  
We can apply the ``cluster'' construction 
(see Sec.~\ref{sec:cluster})
to these cages by representing its lattice Hamiltonian as a sum of cuboctahedron
Hamiltonians, with $j_1=J_1/2$, $j_2=J_2/2$, $j_2=J_2'/2$, and $j_4=J_4/2$
($J_1$, $J_2$, $J_2'$ and $J_4$ are shared by two cuboctahedra); also $j_3=J_3$ and $j_4'=J_4'$.

Consider for a moment the ground state of 12 spins placed
on an isolated cuboctahedron,~\cite{cuboctahedron}
 with couplings $j_1$, $j_2$, $j_3$, and $j_4$.  
Now, it is well known that, on a chain (i.e. a discretized circle),
$j_1>0$ and (small) $j_2<0$ 
give a gradual, coplanar spin spiral; on a circle with the right number of 
sites the spin directions point radial to that circle.
Roughly speaking, the three-dimensional analog of this 
happens on a cuboctahedron, which is a discretized sphere:
if $j_1$ is ferromagnetic and one of the more distant 
couplings is antiferromagnetic, the spin ground state is a 
direct image of the center-to-vertex vector in the cuboctahedron
(modulo a global $O(3)$ rotation of the spins).
This notion only works for the 60$^\circ$ kind of cuboctahedral,
occurring for $J_1>0$, in which the nearest neighbors are 
(relatively) close to being parallel.

To build a global state in which every
cuboctahedron has the cuboctahedral spin configuration, 
the difficult part is just to make the spins agree where they 
are shared between cuboctahedra: that is achieved
by applying mirror operations in alternate layers of 
cuboctahedra, such that e.g. the $s_i^x$ components are multiplied
by $(-1)^{x/a}$.  

\subsection{As special case of $J_1$-only antiferromagnet}
\label{sec:cuboc-J1only}

When we have only $J_1<0$ couplings, the ground state
is a degenerate state with $120^\circ$ angles between 
nearest neighbors, as written in
Eq.~\eqr{eq:J1cuboc} of Sec.~\ref{sec:onlyJ1}.
In that degenerate state of Sec.~\ref{sec:onlyJ1}
the spins take (some of or all of) the cuboctahedral directions, 
but do not have a genuine cubic spin symmetry.
As soon as an arbitrarily small antiferromagnetic $J_2$ 
is added as a perturbation, a subset of these states
is selected, which is the $120^\circ$ kind of 
cuboctahedral state.  In the notation of \eqr{eq:J1cuboc},
this true 120$\circ$ cuboctahedral state 
takes the following form:
   \begin{subequations}
   \begin{eqnarray}
        \ss(n_1+\half,n_2,n_3) &=& 
              \frac{1}{\sqrt{2}} [0, (-1)^{n_2}, (-1)^{n_3}]; \\
        \ss(n_1,n_2+\half,n_3) &=& 
              \frac{1}{\sqrt{2}} [(-1)^{n_1}, 0, \pm(-1)^{n_3}]; \\
        \ss(n_1, n_2, n_3+\half) &=& 
              \pm \frac{1}{\sqrt{2}} [(-1)^{n_1}, (-1)^{n_2},0].
   \end{eqnarray}
   \label{eq:fullcuboc}
   \end{subequations}
where $+$ corresponds to the 120$^\circ$ state and $-$ to the 60$^\circ$.

\subsection{Luttinger-Tisza approach to cuboctahedral states}
\label{sec:cuboc-LT}

Alternatively, both cuboctahedral states can be understood within the LT framework.
For certain domains of parameter values, the optimal LT modes
have wavevectors $\kkLT $ of $\{1/2, 0, 0 \}$ type.
It can be worked out that for e.g. $\kkLT=(1/2,0,0)$, 
one eigenmode has amplitudes $(1,0,0)$ on sublattices $(x,y,z)$, 
i.e. its support is only on the $x$ sublattice. 
Each of the other two eigenmodes has 
its support equally on the $y$ and $z$ sublattices, the
amplitudes being $(0,1,\pm 1)$.
When the first eigenmode is optimal, we get the decoupled
$(1/2,0,0)$ antiferromagnet already described in Sec.~\ref{sec:coplane};
when either of the two-sublattice eigenmodes is stable, 
a cuboctahedral states is found.

It is obviously impossible to satisfy normalization 
in a spin state using just one of the two-sublattice modes, 
since its amplitude vanishes on the third sublattice.  
To build a normalized ground state, it is necessary 
and sufficient to form a linear combination using all 
three of the symmetry-related $\kkLT$ wavevectors, 
associating each with a different orthogonal spin component.
Thus the spin directions are $(1,1,0)/\sqrt{2}$,
with all possible permutations and sign changes,
as we already saw in Eq.~\eqr{eq:fullcuboc}.
These states could be called a commensurate triple-Q state.
The eigenmode with amplitudes of form $(0,1,1)$ gives
the 60$^\circ$ cuboctahedral whereas the one of form
$(0,1,-1)$ gives the 120$^\circ$ cuboctahedral state.

\subsubsection{Absence of (1/2, 1/2, 0) Cuboctahedral State}
\label{sec:cuboc-no}

From the LT viewpoint, one would naively expect to construct 
similar noncoplanar cuboctahedral states of cubic symmetry
using $\kkLT$ of (1/2,1/2,0): why are they absent?
After all, the $\{1/2, 1/2,0\}$ type wavevectors are threefold degenerate, 
just like the (1/2,0,0) wavevectors from which the cuboctahedral states
are built, and it is straightforward to follow the analogy
of those states to construct a (1/2, 1/2,0) cuboctahedral 
(just allotting each mode one of the three cartesian directions
in spin space).
Furthermore, if we include $J_4$ and $J_4'$ couplings in the Hamiltonian,
there is a certain region of parameter space in which $\kkLT$ = (1/2, 1/2, 0)
can indeed be optimal, with the optimal LT eigenmodes being
orthogonal to the eigenmodes that make up the $(1/2, 1/2, 0)$
type antiferromagnetic state.  Hence in that region,
the putative $(1/2,1/2,0)$, cuboctahedral state really is a ground state.

But closer examination of the LT matrix for $\kk=(1/2, 1/2, 0)$
shows that neither the cuboctahedral state, nor any noncollinear state,
is {\it forced}.
At this high symmetry point in the Brillouin zone, all intersublattice 
contributions to the LT matrix $\tJ_{\alpha\beta}(\kvec)$ 
[see Eq.~\eqr{eq:Ham-fourier}] cancel.
That means the $(\tJ_{\alpha\beta})$ is a diagonal matrix.
Its eigenvalues are $-J_2+2J_2'-4J_4-2J_4'$ for the 
mode of the (1/2,1/2,0) antiferromagnet, plus two degenerate 
eigenvalues $J_2+2J_4'$ for the modes of interest here.

Furthermore, the fact that only $J_2$ and $J_4'$ enter the formula
indicates that all other couplings cancel out.  Not only are spins 
of different sublattices  decoupled, but each sublattice decouples 
into two interpenetrating (and unfrustrated) tetragonal lattices .
(The latter decoupling is reminiscent of the decoupling of the 
$J_2$-only simple cubic antiferromagnet.)
In light of these decouplings, we cannot call this state a (1/2, 1/2, 0) 
cuboctahedral; it is merely a particular configuration out of
a degenerate family that also includes collinear states.

\OMIT{
\subsection{Cuboctahedral state as a spiral?}
A fourth way to view the cuboctahedral state is
as a special (commensurate) case of the
transverse-modulated alternating conic spiral, 
where ``transversely modulated'' is elaborated
in Sec.~\ref{sec:transverse-mod} (below).
Such a state consists of a stacking of layers, 
periodic within each layer, and each layer
has the same spin state except for a rotation
about the cone axis $\Chat$.
Consider, e.g., a 60$^\circ$ cuboctahedral,
focusing on one cubic cell around which the spin
directions all point outward from the cell center,
as constructed in Sec.~\ref{sec:cuboc-cages}.
Take the stacking direction to be $z$.
Then $\Chat = \zhat$; the z components alternate
from layer to layer as 
$0,+1/\sqrt{2},0, -1/\sqrt{2},\ldots$, where 
the zeroes occur in the $z$-sublattice and
the repeat is two lattice constants.
(NOTE: The components normal to $\zhat$ don't
spiral at all, they stack periodically.
Well, $Q=0$ is a special case of a spiral,
but that is stretching it.)
}


\OMIT{
Any of the three $\langle 100 \rangle$ directions
may be considered as the stacking direction.
Whenever the $\{1/2,0,0\}$ LT optimum goes unstable by
bifurcating into $\{ 1/2 \pm \delta, 0, 0\}$, we
conjecture a continuous transition of the cuboctahedral
state to the incommensurate version of this
transverse-modulated alternating-conic state,
with one of the $\langle 100 \rangle$ directions selected as
the unique stacking.  (However, this transition need not
happen at exactly the same parameter values that the LT optimum
bifurcates, since the incommensurate state is no longer
built from LT optimum wavevectors.)
}

\OMIT{
Finally, the cuboctahedral state also appears to
be a limiting case of the double-twist kind of state.
(Sec.~\ref{sec:doubletwist}).
}

\section{Conic spiral states}
\label{sec:conic}

The conic spiral states are generically incommensurate and constitute
the most common class of non-coplanar state that we found.
They are layered states, where the spins are all parallel in a given 
layer. That is, the lattice breaks up into layers, normal to
some stacking direction $\Qstack$ in real space. 
We encountered only $\Qstack=\{Q00\}$ stacked conic spirals, so we
concentrate on that case, but stacking directions 
other than $\{100\}$ should be feasible in principle.
Within each layer, all the spin directions are the
same; as you look in each successive layer, the spin
directions rotate around an axis $\hatcone$ in spin space.
Due to this layering, it is possible to map a conic state 
to a  one-dimensional ``chain lattice''
(as introduced in Sec. \ref{sec:mapping}), 
which is a significant simplfication in the analysis.

Considered from the LT viewpoint, a conic spiral is a mix of two 
different modes: two spin components follow an incommensurate
wavevector $\QQ$, spiraling as in a helimagnet, and the third
component follows a commensurate wavevector $\QQ'$
($\QQ'\cong 0$[AF] in the asymmetric-conic case).  In both $\QQ$ and
$\QQ'$, all components transverse to the stacking direction must be zero.

\subsection{Categories of conic spiral states}

The conic spirals divide into subclasses, the alternating and 
asymmetric conic spirals, according to whether they
are symmetric under the (spin space) symmetry of
reflecting in the plane normal to $\hatcone$. 
In the octahedral lattice, the only type of conic spiral we observed was the asymmetric conic. 
However, in the chain lattice we find that two distinct classes of conic spiral are possible: 
the asymmetric conic and the alternating conic.
In principle, both states should be possible in the octahedral lattice,
but an analysis in terms of stacking vectors (see Sec. \ref{sec:mapping}) reveals that longer range couplings ($J_5,J_6,\dots$) would be necessary to stabilize the conics.

\subsubsection{Asymmetric conic spiral}

In the asymmetric conic, the ground states are linear combinations of helimagnetic and ferromagnetic modes. 
Let the stacking direction be along the $z$ axis, so all spins are only functions of $z$.
(Alternatively, we could interpret $z$ as the position in the one-dimensional chain
lattice of Sec.~\ref{sec:mapping}, below.)
The helimagnetic part is parametrized with a rotating unit vector:
   \begin{equation}
      \Ahat(z) \equiv \cos(Q z) \Ahat_0 + \sin(Q z) \Bhat_0
   \label{eq:rotating-Ahat}
   \end{equation}
where $(\Ahat_0,\Bhat_0,\Chat)$ form an orthonormal triad.
Then
   \begin{subequations}
   \begin{eqnarray}
        \ss(z) &=& 
              \cos(\alpha) \Ahat(z) + \sin(\alpha) \Chat; \\
        \ss(z+\half) &=& 
              \cos \beta   \Ahat(z+\half)  - \sin(\beta) \Chat,
   \label{eq:asymconic}
   \end{eqnarray}
   \end{subequations}
with $0 \le \alpha,\beta \le \pi/2$.
The spins in each sublattice rotate about some common cone axis $\Chat$ in spin space;
spins of the same sublattice have the same component along the cone axis 
(giving net magnetic moments for both sublattices of $N*\ss_\Vert^{(\alpha)}$, where $\ss_\Vert^{(\alpha)}$ is the component of a spin of sublattice $\alpha$ along the common axis).
The different sublattices are antiferromagnetically coupled, so these net moments have different signs. And because the couplings within the sublattices are not equal to those of the other sublattice, the magnitude of the net moments are not equal. This can be easier to understand if we think in terms of common-origin plots (see Figure~\ref{fig:cop}) , since then the spins all lie upon the surface of a sphere. In the common origin plot, each sublattice forms a cone. These cones are along the same axis, but oppositely oriented and their azimuthal (conic) angles are not equal. 

\subsubsection{Alternating conic spiral}

In the alternating conic, one sublattice is a {\it planar} spiral, 
while the other is always a combination of the same helimagnetic mode 
and an antiferromagnetic mode.  
This state is thus represented, again using \eqr{eq:rotating-Ahat}:
   \begin{subequations}
   \begin{eqnarray}
        \ss(z) &=& 
              \cos(\alpha) \Ahat(z) + (-1)^{z}\sin(\alpha)\Chat; \\
        \ss(z+\half) &=&  \Ahat(z+\half);
   \label{eq:altconic}
   \end{eqnarray}
   \end{subequations}
where $0 \le \alpha \le \pi/2$ for the alternating conic.
Returning again to our common-origin plot (see Figure~\ref{fig:cop}), one sublattice forms a great circle along the equator of the sphere. The other sublattice now traces cones on each side of this equator (the common axis is the vector normal to the circle). The spins of the second sublattice alternate between the two sides of the equator, giving the antiferromagnetic component. Thus, the difference between sublattices is more fundamental in the case the alternating conics than in asymmetric conics. 

\subsubsection{Splayed States}
There are two other states essentially related to these conic spirals, but are important enough to deserve their own names (this is much the same as ferromagnetism being a special case of helimagnetism). We term these states ferromagnetic and ferrimagnetic splayed states. 

Consider the alternating conic in the limit of the polar angle going to 0 or 1/2. In both cases, the spins are confined to a plane, but they are emphatically not in a helimagnetic configuration. The sublattice that was helimagnetic is now ferromagnetic and the sublattice that was conic now merely alternates (that is, reflects about equatorial plane without rotation). If the polar angle is 0 (1/2), then the dot product of spins in different sublattice is positive (negative) and the state is a ferromagnetic (ferrimagnetic) splayed state.

While the difference between ferromagnetic and ferrimagnetic splayed state seems rather trivial here, it is more dramatic when we think of asymmetric conics. The ferrimagnetic splayed state is produced when one of the conic angles and the polar angle go to 1/2 while the other conic angle remains arbitrary. But because the coupling between sublattices is antiferromagnetic for the asymmetric conic, the conic angles of the sublattices will always confine spins to opposite sides of the "equator." This means that the asymmetric conic will never continuously transform into a ferromagnetic splayed state, and so such a transition would necessarily be first order.

\subsection{Stackings and chain mapping}
\label{sec:mapping}

For the cases of incommensurate conic spirals, our main analytic 
method is variational: we assume a functional form for the spin 
configuration (based on interative minimization results) depending 
on several parameters, and optimize exactly with respect to them.  
Say we know that the correct ground state is stacked
a stack of planes with identical spins --
-- in practice this is determined empirically from
the outcome of iterative minimization (Sec.~\ref{sec:iter-min}) --
then the variational problem is equivalent to a one-dimensional
(and hence simpler) one: layers of the 3D lattice may be mapped 
into a chain containing inequivalent sublattices.

This mapping is fruitful in two or three ways.  First, we could 
(and did) explore the chain lattice ground states using interative 
minimization, in much longer system lengths than would be practical
in an $L\times L \times L$ system. Second, it is unifying, in that
various stackings of various three-dimensional lattices map to the 
same chain lattice. Finally, it illuminates what conditions are
necessary in order to obtain non-coplanar states.  

Notice that if
a stacked state is the true ground state of the three-dimensional
lattice, its projection must be the true ground state of the
chain projection (since the chain lattice states correspond exactly
to a subset of three-dimensional states); but of course the converse 
is false:  the proven optimal state of the chain lattice might be
irrelevant to the three-dimensional lattice, when a different 
(e.g. unstacked) kind of ground state develops a lower energy.
As coupling parameters are varied, that different state might become
stable in a first-order transition; our only systematic ways to
address that possibility are (i) iterative minimization (ii)
watching for an exchange of stability between two LT eigenmodes
at different wavevectors. And even with this method, 
we had to rely primarily upon iterative minimization 
for reliable results, as LT analysis is insufficient
to determine the ground state, particularly in cases
where a ground state cannot be constructed from the optimal 
LT modes. 

For the cases that concern us here, the chain lattice has a basis
of two sites per cell, with inversion symmetry at each site;
we take the lattice constant to be unity.
The mappings to chain sites $z$ is given by
    \begin{equation}
       z = \Qstack \cdot\rr
    \end{equation}
where $\Qstack$ is a vector of integers, having no common factor.
We let ``even'' sites be those with $z$ integer and ``odd''
sites be those with $z=$ integer+1/2.  
As in three dimensions, we consider inter-sublattice couplings
$j_1$ and $j_3$, as well as intra-sublattice couplings
out to distances 1 and 2, namely $j_2$ and $j_4$ (between even spins)
or $j_2'$ and $j_4'$ (between odd spins).   
Notice that, if $j_2=j_2'$ and $j_4=j_4'$, the chain system 
reduces to a Bravais lattice (with lattice constant 1/2) and
its ground states are (at most) coplanar spirals, as explained
in  Sec.~\ref{sec:LT}; that rules out $\Qstack=(1,1,1)$.
Notice that for stackings in low symmetry directions
(and thus requiring larger coefficients in the $\Qstack$ vector),
short range $J_{ij}$ couplings in the octahedral lattice map
to long range $j_{ij}$ couplings in the chain lattice, e.g.
$\Qstack=(211)$ maps $J_{1}$ through $J_{4}$ to 
$j_{1}$ through $j_{6}$. Because longer range couplings
quickly appear, it is reasonable to explore them in the chain lattice.
In order to organize our exploration of parameter space, we shall
call $j_1$, $j_2$, and $j_2$ ``primary'' couplings; 
$j_3$, $j_4$, and $j_4'$ are ``secondary'' couplings, and if
necessary are assumed small compared to the primary couplings.

\begin{figure}
\includegraphics[scale=0.6]{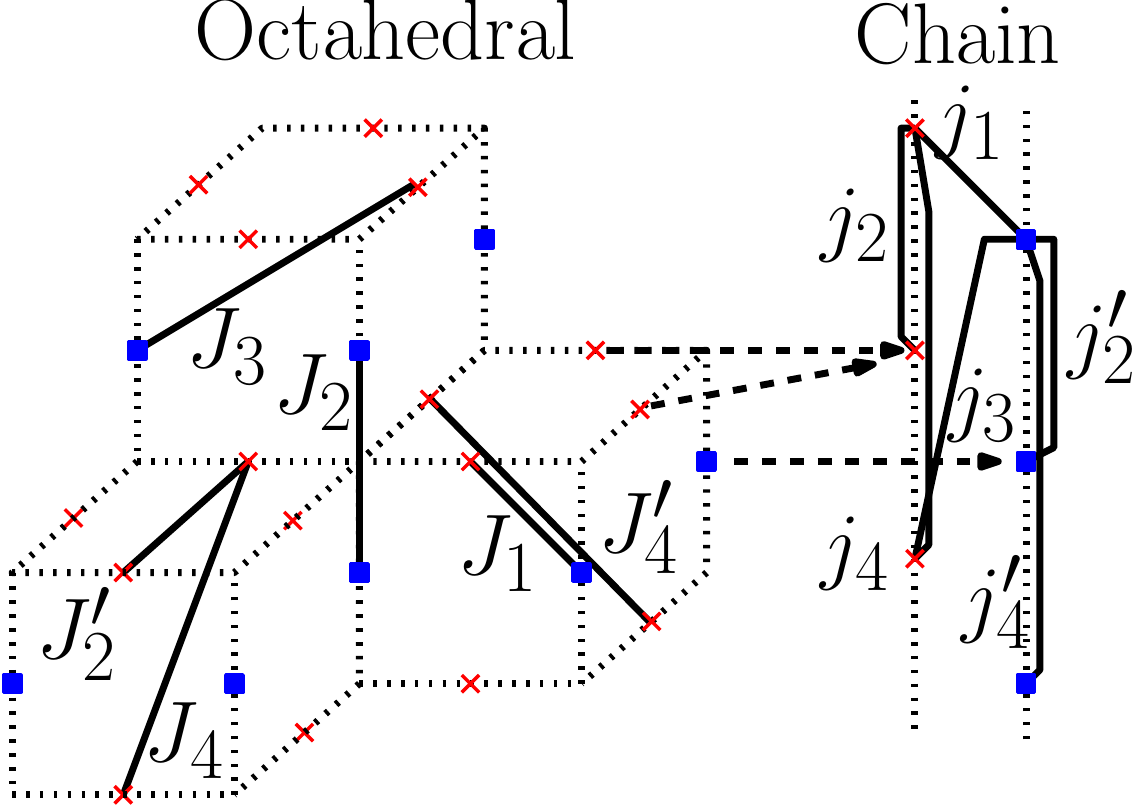}
\caption
{(COLOR ONLINE) 
The mapping between the octahedral and chain lattices for a (001)
stacking direction. All sites with the same value of z on the $x$
and $y$ sublattices (red x's) are projected onto the same
point, while those in the $z$ sublattice (blue squares) are
mapped onto a different point (again a function z). As such, there
are twice as sites mapped onto any x as there are mapped
onto a square. The spatial structure is shown in dotted lines,
while couplings are shown in solid lines.}
\label{fig:mapping}
\end{figure}

We encountered $\Qstack=(1,0,0)$ stackings often enough in the iterative
minimization, and we searched for $\Qstack=(1,1,0)$ type stackings 
also  (although this search was ultimately unsuccessful).
The $\Qstack=(001)$ mapping is illustrated in real space in 
Fig.~\ref{fig:mapping}.  The numerical values of the mapped
couplings is given by a matrix multiplication:
   \begin{equation}
		\label{eq:mapping}
  \begin{pmatrix}
     j_{0}\\
	  j_{1}\\
     j_{2}\\
     j_{2}'\\
     j_{3}\\
     j_{4}\\
     j_{4}'
\end{pmatrix}=
\begin{pmatrix}
4 & 2 & 4 &  0& 4 & 2\\
8 & 0 & 0 & 16 & 0 & 0 \\
0 & 0 & 2 & 8 & 4 & 4\\
0 & 1 & 0 & 0 & 4 & 0\\
0 & 0 & 0 & 0 & 0 & 0\\
0 & 0 & 0 & 0 & 0 & 0\\
0 & 0 & 0 & 0 & 0 & 0
\end{pmatrix}
\begin{pmatrix}
  J_{1}\\
  J_{2}\\
  J_{2}' \\
  J_{3}\\
  J_{4}\\
  J_{4}' \\
\end{pmatrix}
\end{equation}
where $j_{0}$ is the energy within a plane of constant $z$ in the octahedral lattice. Notice that, for this stacking vector, couplings $J_1$ through $J_4$ are projected down to $j_0$ through $j_2$. This helps explain the absence of stable conics in the octahedral lattice. We only find conics in the chain lattice for $j_3$ or longer-range couplings. But for this stacking $\Qstack=(001)$, require at least a $J_5$ coupling to generate an analogous coupling (for the alternating conic, $J_6$ is more likely, given the asymmetry between the sublattices).

\subsection{Transversely modulated spirals}
\label{sec:transverse-mod}

This is a hypothetical (but likely) class of states.
Here ``transversely modulated'' means that when we
decompose the lattice as a stacking of layers,
a single layer does not have a single spin direction,
but instead a pattern of spin directions.
Whereas the asymmetric conic spiral used ordering 
wavevectors (say) $(Q,0,0)$ and $(0,0,0)$, 
and the alternating conic spiral used $(Q,0,0)$ and
$(1/2, 0,0)$, a transversely modulated spiral
might replace the first wavevector by e.g.
$(Q, 1/2, 1/2)$.

Equivalently, if we look at a column of
successive cells along the stacking direction, in
a plain conic spiral (whether alternating or asymmetric), 
adjacent columns are in phase, but in the transversely modulated
conic spirals, different columns are offset in phase according
to a regular pattern.  It should be possible to generalize
the chain mapping to such cases, but we have not tried it.

\section{Double-twist state}
\label{sec:doubletwist}

Here we describe the incompletely understood ``double-twist'' state, 
which has attributes in common with both cuboctahedral and conic states,
and was observed only for a small set of couplings 
($J_1=-$2, $J_2<0$, $J_3$=1, all others zero). 
These couplings were selected to give $\kkLT=(Q,Q,0)$, 
as the ground states previously encountered had a $\kkLT$ either along the 
$(111)$ direction or on the edges of the Brillouin zone. 
For a given $Q$,  the value of $J_2$ is determined by:
\begin{equation}
	J_2 = 4\sqrt{2}\cdot(3\cos^2(Q/2)-1)/\sqrt{1-2\cos^2(Q/2)}.
\label{eq:J2-Q-doubletwist}
\end{equation}
We particularly studied the $Q=3/8$ case.
(Note that iterative minimization necessarily probes commensurate states,
due to our boundary conditions.)
This corresponds to $J_2/J_3=-3.7717$, according to
Eq.~\eqr{eq:J2-Q-doubletwist}.

The double twist state is, to good approximation, composed solely of 
$(Q,Q,0)$ modes (for normalization, there will necessarily be other 
wave-vectors, but these have relatively small amplitudes).
Unlike previous states, each sublattice has nonzero contributions from
all $(Q,Q,0)$ wavevectors, rather than a subset. 
The weight of each sublattice in a given $(Q,Q,0)$ mode differs between sublattices,
approximately in proportion to relative weight in the LT optimal mode with 
a similar $\kkLT$.

The spatial variation produced by this combination of modes is complicated. 
There is a stacking axis in real space, which we take to be
$\hat z$ without loss of generality.
Spin space is characterized by three orthonormal basis vectors:
$\Chat$ defines a conic axis, around which the other 
two basis vectors $\Ahat$ and $\Bhat$ rotate as a function of $z$:
  \begin{subequations}
  \label{eq:A-B-basis}
  \begin{eqnarray}
     \Ahat(z)&=& \cos(Qz)\Ahat_0 -\sin(Qz)\Bhat_0; \\
     \Bhat(z)&=&  \sin(Qz)\Ahat_0 + \cos(Qz)\Bhat_0; \\
     \Chat &=& \Ahat(z)\times \Bhat(z) = \Ahat_0\times \Bhat_0.
  \end{eqnarray}
  \end{subequations}

We can parameterize this cartoon of the double twist state  as
\begin{equation}
\begin{pmatrix}
     S_1(\rr)\\ S_2(\rr)\\ S_3(\rr)\\
\end{pmatrix}
= \Gamma_0
\begin{pmatrix}
-a & b & b \\
b & -a & b \\
b & b & -a \\
\end{pmatrix} 
\begin{pmatrix}
\sin(\Phi_y)\Ahat(z) \\
\sin(\Phi_x)\Bhat(z) \\
\ \Gamma \cos(\Phi_x)\cos(\Phi_y)\Chat \\
\end{pmatrix} 
    \label{eq:doubletwist}
    \end{equation}
with $\Phi_x(c)\equiv Qx-\phi_x$, where $\phi_x$ is an arbitrary phase, 
similarly $\Phi_y(y) \equiv Qy-\phi_y$.
Note the coordinates $\rr$ in $\SS_i(\rr)$
are the actual sites for sublattice $i$, which are half-odd-integers
in the $i$ component.
Thus, addition to the twisting of the basis vectors along the stacking direction,
in \eqr{eq:A-B-basis},
there are spatial modulations transverse to the stacking direction that 
appear in the coefficients of $\Ahat,\Bhat$ and $\Chat$ in 
\eqr{eq:doubletwist}
\OMIT{This modulation is always of the form $\cos(Q\cdot r_i-\phi_i)$, 
where $r_i$ is one of the two vectors orthogonal to the axis rotation vector. 
Each of the two rotating axes is modulated by an orthogonal $r_i$, 
while the conic axis is modulated by both of these factors.
The maximum amplitude along each of basis vectors varies by sublattice 
in proportion to the amplitude of different LT eigenmodes.}

While the form described by \eqr{eq:doubletwist} is close to what we observe with 
iterative minimization, it unfortunately does not satisfy normalization:
the ground state necessarily contains admixtures of non-optimal modes. 
(To satisfy normalization using only the $(Q,Q,0)$ modes would 
require four-component spins.) 
In Eq.~\eqr{eq:doubletwist},
$(-a,b,b)$ should ideally be the amplitudes (on the three respective 
sublattices) of the LT eigenvector at $(0,Q,Q)$, while
$\Gamma$ is a weighting factor that reduces the deviations of
the spins in \eqr{eq:doubletwist} from uniform normalization.

What if we demanded, not normalization of all spins,
but only that the mean-squared value of 
$|\SS_i(\rr)|^2$ be one in each sublattice? 
Since each cosine or
sine factor has mean square of 1/2, and since $a^2+2b^2=1$,
we ought then to have $\Gamma_0=\Gamma=\sqrt{2}$.
Projecting the actual result of iterative minimization onto such modes 
gave $\Gamma \approx 1.36$.  Also, whereas
$b/a=1.64$ in the actual LT eigenvectors,
we found $b/a \approx 1.5$ in the results of iterative minimization.
\OMIT{It is worth emphasizing that this state is invariant under a global rotation, 
so the specific x, y, z variations given here are not intrinsic to the state.}

The double twist state can be viewed as related to the hypothesized transversely-modulated conics or to the cuboctahedral states. In particular, the composition of this state in terms of LT modes is more similar to the cuboctahedral states than to any other configuration.

The LT mode underlying this state, according to
\eqr{eq:J2-Q-doubletwist}, has a continuously variable
wavevector as $J_2$ is varied. Due to the limitations of
iterative minimization with periodic boundary conditions,
we have not followed the evolution of the double-twist
state; in particular, we do not know if it becomes 
incommensurate in both the stacking ($z$) direction
and the transverse directions.

\SAVE{CLH had imagined that the double-twist state can be viewed as
an ideal LT state using 4 spin components, that has been 
projected down to 3 spin components.  However, at best that
would work using an eigenvector of form $(\pm a, \pm a, \pm b)$,
which is not one of the LT modes.}

\section{Phase diagrams}
\label{sec:phase}

To understand how the ground states outlined in Sec.~\ref{sec:coplane},\ref{sec:cuboc}, \ref{sec:conic}, and \ref{sec:doubletwist} fit together, it is necessary to examine the phase diagrams of the octahedral and chain lattices. 
A series of representative cuts through the phase diagrams for 
both lattices gives us a general sense of their topology, and 
specifically in what regions non-coplanar states are stabilized.

Of course, rescaling the couplings by any positive factor
gives an identical ground state (with energy rescaled by the
same factor).
Therefore we present the phase diagrams in rescaled coordinates, 
normally $J_i \to J_i/|J_1|$ (except when $J_1=0$).

An important aspect of all the phase diagrams is the classification
of the transitions into first-order (discontinuous),
encompassed (continuous), or degenerate: the distinction between
the last two kinds was explained in Sec.~\ref{sec:bridging}.
Whenever a continuous manifold of degenerate states is found 
(always on a phase boundary), it is labeled in the diagrams
by ``$O(L^d)$'' representing how the number of parameters 
(needed to label the states) scales with system size.

\subsection{Octahedral Lattice}

In the octahedral lattice we are fortunate in that most
kinds of states have energies that can be written exactly
as a linear combination of couplings
(given in Table~\ref{tab:O-States}~\cite{FN-asymm-conic});
the phase boundaries between such phases are simply
the lines (more exactly hyperplanes) where the
two energy functions are equal.
Most other phase boundaries 
are handled analytically, e.g. the helimagnetic state and its
``encompassed'' ferromagnetic and antiferromagnetic states.
The only phase boundary {\it not} determined analytically
from a variational form was the double-twist state, for which
we do not have an exact variational form; in this case
the boundary was approximation by the LT phase boundary.
We would expect that approximation to be accurate for any 
such complex phase that is built entirely from a star of 
symmetry-equivalent modes, provided the neighboring phase 
is built from other modes.


\begin{table}
\begin{tabular}{|c|c|c|}
\hline 
State&  $\QQ$ & Energy/spin\tabularnewline
\hline
\hline 
---
Ferromagnetic  & (000)  & $-4J_X-J_2-2J_2'-4J_4-2J_4',$
\tabularnewline
 & & $J_X \equiv J_1+2J_3$ \tabularnewline
\hline
3 sublattice - 120$^\circ$ & (000) & $2J_X-J_2-2J_2'-4J_4-2J_4'$\tabularnewline
\hline 
(1/2,1/2,0)-AFM  & $(\half,\half,0)$ & $-J_2+2J_2'-4J_4-2J_4'$\tabularnewline
\hline 
(1/2,1/2,1/2)-AFM & $(\half,\half,\half)$ & $J_2+2J_2'-4J_4-2J_4'$\tabularnewline
\hline 
(1/2,0,0)-AFM & $(\half,0,0)$ & $J_2-2J_2'-4J_4-2J_4'$\tabularnewline
\hline 
$\pi$/3-Cuboctahedral & $(\half,0,0)$ & $-2J_1-J_2+4J_3+4J_4$\tabularnewline
\hline 
$2\pi/3$-Cuboctahedral & $(\half,0,0)$ & $2J_1-J_2-4J_3+4J_4$\tabularnewline
\hline 
Helimagnet  & $(qqq)$ & $-2J_1-2J_4-J_4'-J_S^2/(8J_L),
$\tabularnewline
&  & $J_S \equiv 2J_1+J_2+2J_2'+4J_3$ \tabularnewline
&  &  and $J_L \equiv 2J_3+2J_4+J_4'$ \tabularnewline
\hline
\end{tabular}
\par
\caption{Ground States of the Octahedral Lattice,
with ordering wavevector $\QQ$.}
\label{tab:O-States}
\end{table}

Using this information, we can easily find the phase boundaries of various states. 
To aid in graphical display, we will normalize all couplings by $|J_1|$ and 
restrict attention to $|J_3|/|J_1|<1$.  Phase diagrams will be plotted in
the variables $(J_2,J_2')$ representing a slice with $(J_1,J_3)$ fixed.
In all such slices,  the second, third, and fourth
quadrants of the phase diagram are dominated by antiferromagnetic phases of
ordering vector $(1/2,1/2,0)$, $(1/2,1/2,1/2)$, and $(1/2,0,0)$ respectively.
Recall that all of these are {\it nontrivially decoupled} states, in which 
distinct sublattices can be independently rotated due to cancellations
of the inter-sublattice interactions.  When thermal or quantum fluctuations 
are added to the description, ``order-by-disorder effects''~\cite{shen82,henley89}
typically select specific states from these manifolds that are {\it collinear}.
The first quadrant is dominated either by the ferromagnetic phase, or (if
$J_1<0$) by the 120$^\circ$ 3 sublattice phase.  Cuboctahedral phases may
be found near the $J_2'=0$ axis when $J_2>0$.

The phase transitions are always first order in the octahedral lattice, with the following exceptions,
which can be classified according to the three scenarios for bridging states outlined in
Sec.~\ref{sec:bridging} (encompassing and degenerate).
\begin{itemize}
\item[(1)]
The transition from the helimagnet to either the (1/2,1/2,1/2) antiferromagnet 
or to the ferromagnet 
is continuous, since the optimal wavevector varies continuously along $(q,q,q)$
until it hits the commensurate value ((1/2,1/2,1/2) or (0,0,0)), then stops;
this is an example of an encompassing state.
\item[(2)]
Transitions between two antiferromagnet phases are always degenerate, since the
phase boundaries in parameter space are given by $J_2$=0 or $J_2'$=0, which (trivially)
decouple sublattices (implying a degenerate family of states).
In the Brillouin zone, the wavevector can evolve continuously along (1/2,q,k) and (1/2,1/2,q) (for the (1/2,0,0) and (1/2,1/2,0) antiferromagnets, respectively).
\item[(3)]
Lastly, transitions between states of the same $\kkLT$ are degenerate,
occurring where two eigenvalues of the LT matrix for $\kkLT$
cross, as a function of changing parameters. 
This is found for the (0,0,0) modes (ferromagnetism and the 120$^\circ$ state) and the (1/2,0,0) modes (both types of cuboctahedral states with each other and with the $\kkLT=(1/2,0,0)$ antiferromagnetic state).
Because the degeneracy is limited to different eigenmodes of the same wavevector, 
these are $O(L^0)$ {\it degenerate} transitions.
\SAVE{=``simply degenerate'' in the sense of Sec.~\ref{sec:bridging}.}
\end{itemize}

\begin{figure}
\includegraphics[scale=0.6]{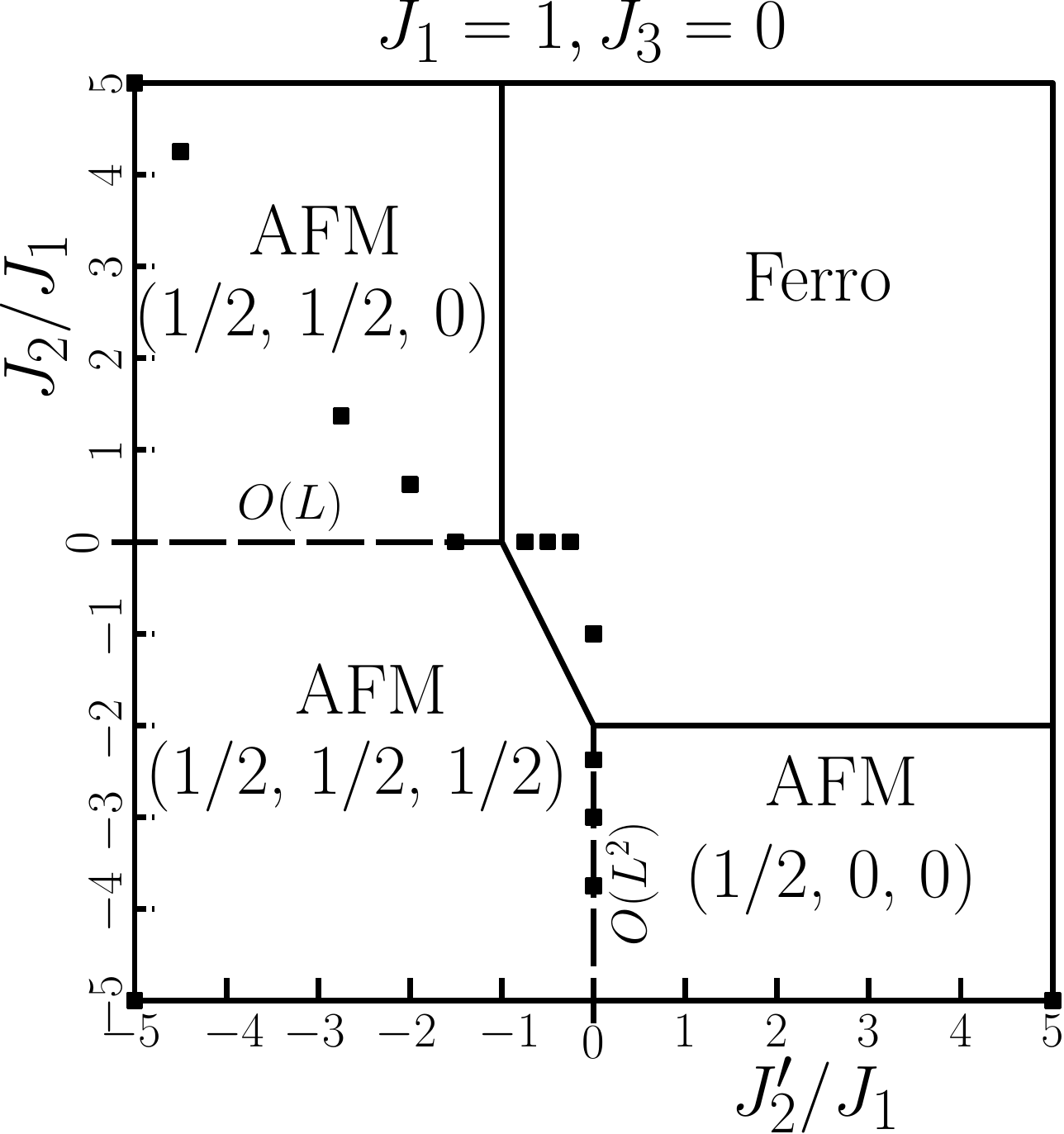}
\caption
{Octahedral Phase Diagram for $J_1=1$, $J_3=0$.
Squares indicate couplings tested with iterative minimization.
\SAVE{(or, for many closely spaced points, dash-dotted lines are used).
(The actual couplings simulated were scaled together so $J_1=1$).}
Solid lines denote first order transitions , dashed lines denote 
degenerate transitions (Sec.~\ref{sec:bridging}), where the scaling of degrees of freedom 
is labeled. Dotted lines indicate a second order transition from encompassing states. 
Regions shaded gray indicate a non-coplanar phase.
}

\label{fig:O,1,0}
\end{figure}

Consider first the phase diagram produced with ferromagnetic $J_1$ and no couplings beyond $J_2,J_2'$ (Figure \ref{fig:O,1,0}). In this case, we find only four states, all of them coplanar (these are outlined in \ref{sec:coplane}).
What's particularly important here, though, is the way that the phase diagram divides up into four quadrants. This is a fairly generic feature that we will see in other phase diagrams.
Along the $J_2'=0$ line between the (1/2,1/2,1/2) AFM and 
the (1/2,0,0) AFM, in all the phase diagrams, we get a degenerate 
($O(L^2)$) decoupled state,
in which each $J_2'$-coupled line has an independent staggered spin direction.

\begin{figure}
\includegraphics[scale=0.6]{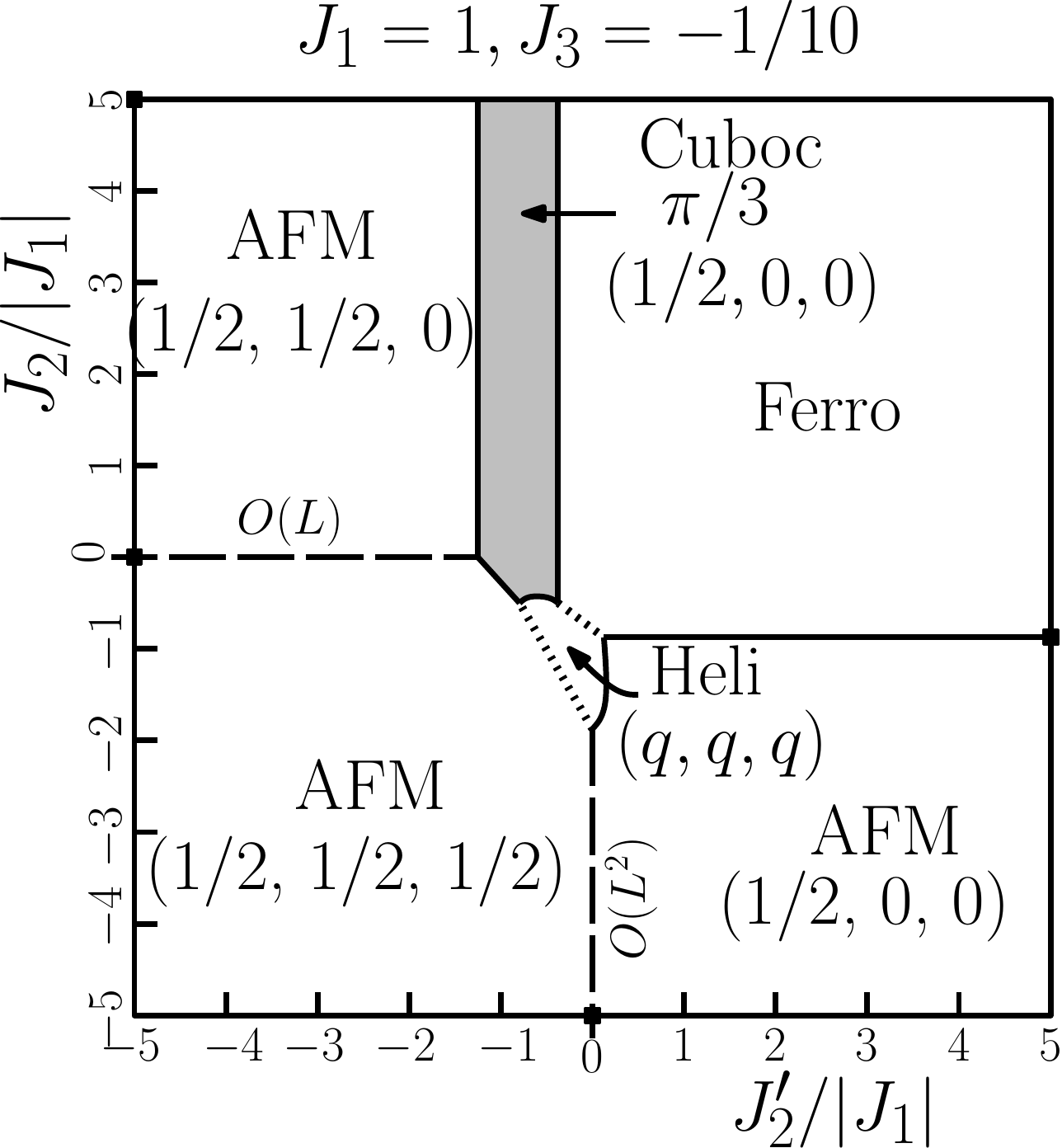}
\caption
{Octahedral Phase Diagram for $J_1$=1, $J_3=-1/10$.
First instance of a non-coplanar state (the $\pi$/3 cuboctahedral
state, in the shaded region) and of helimagnetism.
\SAVE{Here and below, dotted lines indicate second order transition 
by encompassing parametrization (Sec.~\ref{sec:bridging}).}
}
\label{fig:O,1,-1/10}
\end{figure}

Let's now examine how the ferromagnetic $J_1$ phase diagram is modified by
an antiferromagnetic $J_3$ (Figure \ref{fig:O,1,-1/10}). First of all, it stabilizes the $\pi$/3 cuboctahedral state, 
our first example of a non-coplanar phase.
$J_3$ also stabilizes a $(q,q,q)$ helimagnet at the center of the phase diagram.
The boundaries of this phase are quite sensitive to $J_3$:
as $J_3$ becomes more antiferromagnetic, the helimagnet's phase boundaries 
with ferromagnetism and (1/2,1/2,1/2) antiferromagnetism move outward in opposite directions, so as to increase the region of parameter space that is helimagnetic. 
Meanwhile, the phase boundaries of helimagnetism with the (1/2,0,0) antiferromagnet and $\pi$/3 cuboctahedral state move {\it inwards} in opposite directions, so as to decrease the region of parameter space that is helimagnetic. 
The result is that, as $J_3$ becomes more antiferromagnetic, the helimagnetic
region of parameter space first grows and later shrinks until $J_3$=$-J_1/2<0$, where it disappears entirely.

\begin{figure}
\includegraphics[scale=0.6]{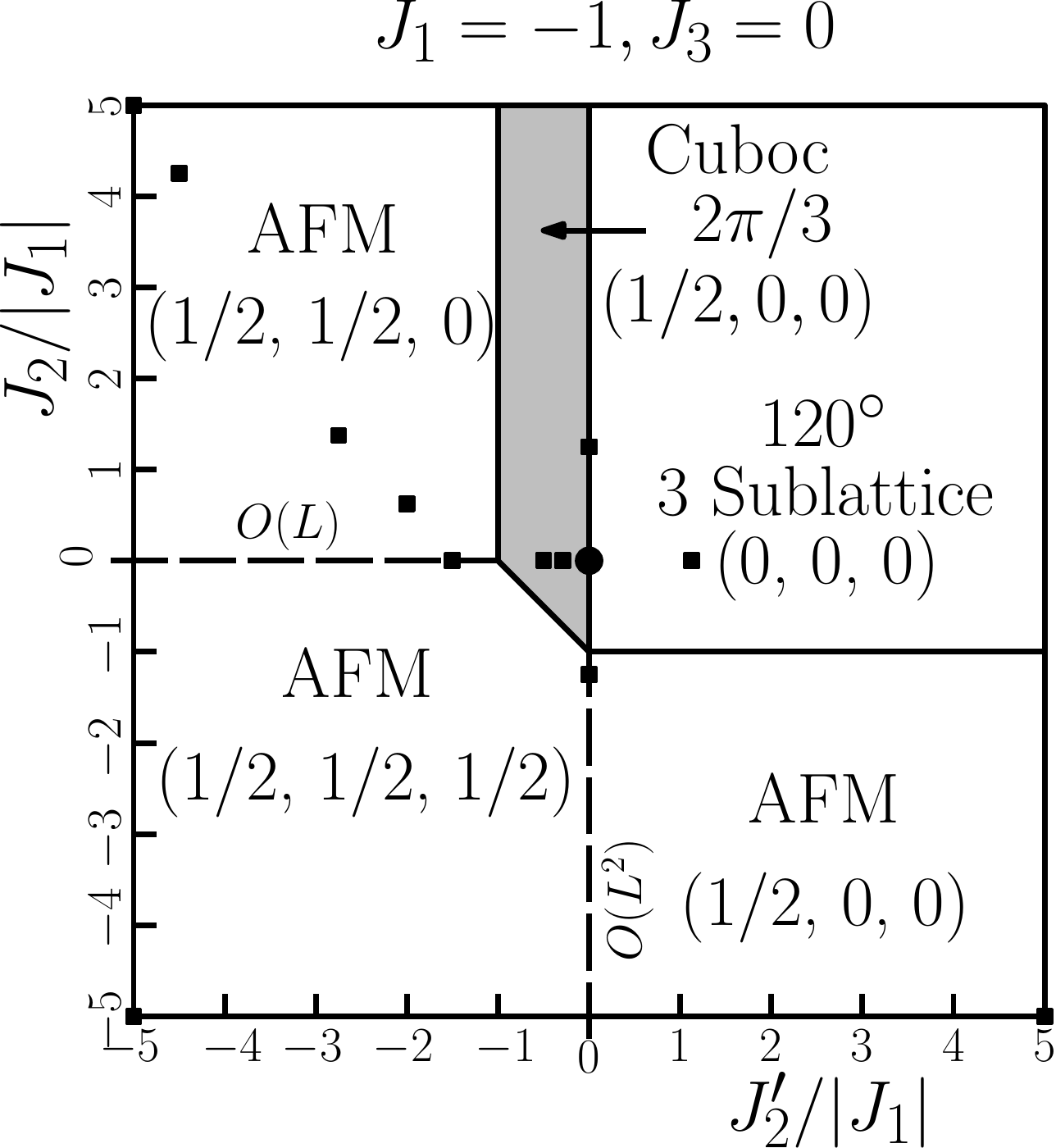}
\caption
{Octahedral Phase Diagram for $J_1=-1$, $J_3$=0.
First instance of the 2$\pi$/3 cuboctahedral state.
The large circle point is the case of only $J_1$ couplings,
corresponding to the (highly degenerate) $120^\circ$ state.
\SAVE{Has been checked that the transition 120$^\circ$-3-subl to 
cuboc($2\pi/3$) is non-degenerate.}
}
\label{fig:O,-1,0}
\end{figure}

Now we turn to the phase diagrams with antiferromagnetic $J_1$,
first considering  arbitrary $J_2,J_2'$ with $J_3=0$
(Figure \ref{fig:O,-1,0}). 
In this case, we still see the quadrant structure, at least qualitatively;
the upper right quadrant now represents the (ordered) three-sublattice 
120$^\circ$ state.
However, a strip between the upper quadrants is occupied by
the 2$\pi$/3 cuboctahedral state, a non-coplanar state 
which only requires two non-zero couplings. 
Furthermore, we find the (highly degenerate) ``$J_1$-only'' state along the boundary between the 2$\pi$/3 cuboctahedral state  and the three-sublattice
120$^\circ$ state.
It is interesting to note that at this point in parameter space, the type of transition changes. 
Because of the vanishing of the couplings, the minimum eigenvalue is a constant along (q,0,0). 
This makes the phase transition at this point a degenerate ($O(L^1)$), 
continuous transition, even though it is elsewhere first order.
\SAVE{SS added on 3-11-12.  I think I would call this a ``degenerate'' not a second order transition.}

\begin{figure}
\includegraphics[scale=0.6]{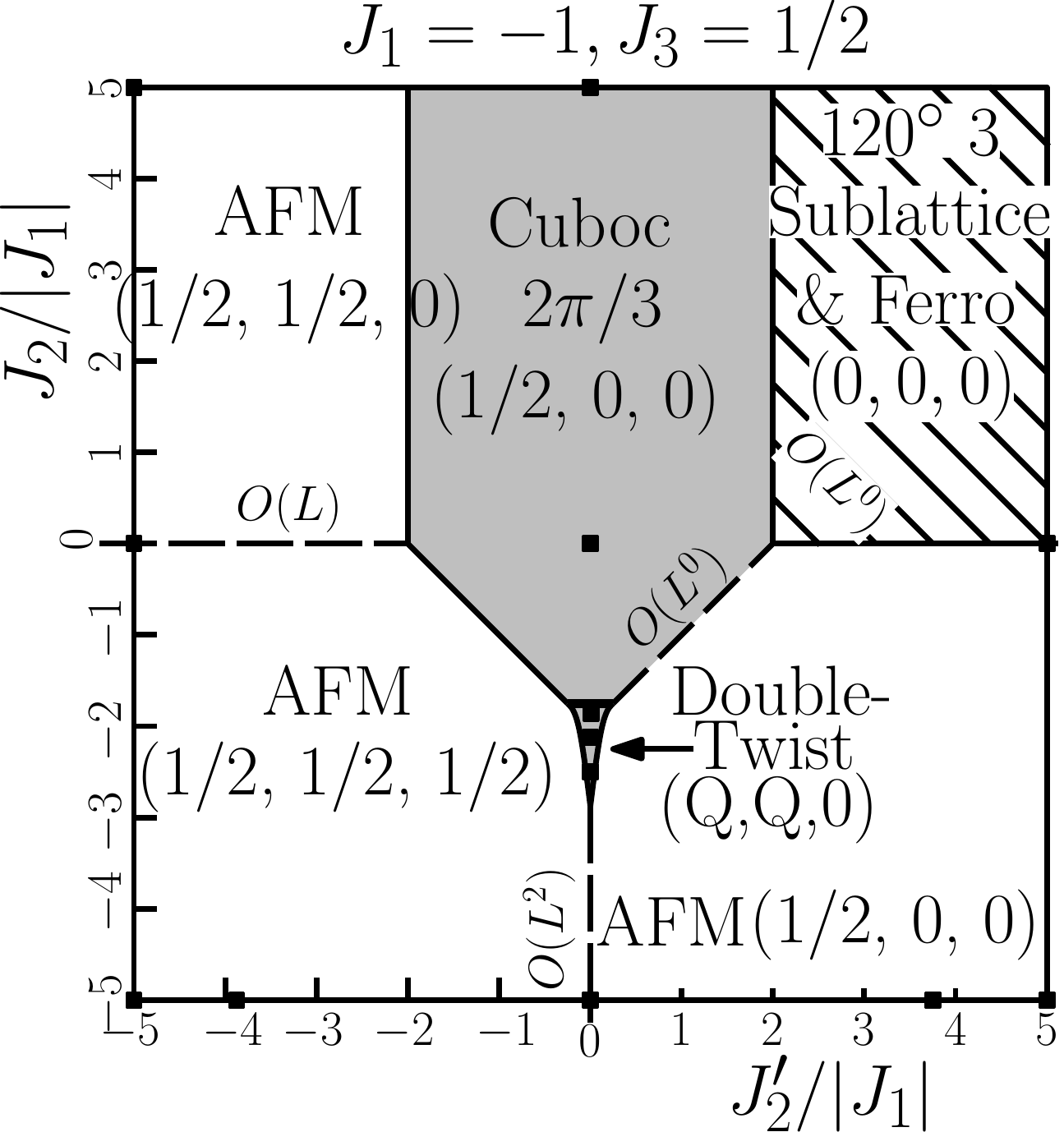}
\caption
{Octahedral Phase Diagram for $J_1=-1$, $J_3$=1/2.
The double-twist state is found in the small region near the origin.
The striped region indicates this slice ($J_3=1/2$) contains a phase
boundary between the 120$^\circ$ 3-sublattice and ferromagnetic states;
for the parameter set plotted, that region has states of the degenerate (marked
``$O(L^0)$'' parameters) and nontrivially decoupled kind.  
If we varied $J_3$ to pass through that region of parameter space,
we would cross a degenerate phase transition.
\SAVE{The boundary between the cuboc domain and the 
combined 120$^\circ$-3-sublattice/ferro domain is non degenerate.
Generically, the LT picture indicates a jump between $Q=0$ and $Q=1/2$, 
and thus a first-order transition.}
}
\label{fig:O,-1,1/3}
\end{figure}

For antiferromagnetic $J_1$ adding ferromagnetic $J_3$ 
(Figure \ref{fig:O,-1,1/3}),
we see the 2$\pi$/3 cuboctahedral state expand, changing the topology of the phase diagram 
(it now shares a boundary with the $(1/2,0,0)$ antiferromagnet). 
In addition, near the triple point of the $(1/2,1/2,1/2)$ antiferromagnet, $(1/2,0,0)$ antiferromagnet 
and 2$\pi$/3 cuboctahedral state, we find the 
double twist state~\cite{FN-doubletwist-figure}.
Note that the slice of parameter space shown here, $J_1=-2J_3$, 
includes the phase boundary between the three-sublattice 120$^\circ$ state 
and the ferromagnetic state, and on the boundary has an extra degeneracy
of the kind described in Sec.~\ref{sec:bridging}.

The double-twist can be understood as a selection from the family of degenerate states. 
Along the $J_2'=0$ line (boundary), we have a degenerate form of decoupled state:
sublattices decouple by row, giving a large 
degeneracy of ground states. 
However, this decoupling depended upon the alternating order within each sublattice
which led to cancellations.
A spiral distortion (selected by adding $J_1,J_3$ to $J_2$ only state, with $J_2$ sufficiently small) within a sublattice allows a non-cancelling interaction 
between it and another sublattice, which lowers the energy.

The $J_1<0$ phase diagrams superficially resemble the
the $J_1>0$ phase diagrams, with two different cuboctahedral 
states appearing around the $J_2>0$ axis, and a helimagnet 
or double-twist state (respectively) appearing in a small
edge below the phase diagram's center.


We have also considered the case of $J_1=0$ with an antiferromagnetic $J_3$ 
(phase diagram not shown).
This phase diagram, apart from the trivial change of normalizing the couplings 
by $|J_3|$ instead of $|J_1|$, strongly resembles the case of antiferromagnetic
$J_1$ and $J_3=0$ shown in Figure \ref{fig:O,-1,0}); the sole difference is
that we now find the $\pi$/3 cuboctahedral state in place of the 2$\pi$/3 cuboctahedral.

\SAVE{At this special point, these conic spirals are presumably
exact LT states, made by mixing inequivalent but ``accidentally''
degenerate modes.  
It would appear that the linear combinations of modes that
satisfy unit-spin normalization are all conic spirals, apart from
the limiting configurations, forming perhaps a one-parameter family?
However we have not tested this analytically; certainly they're what we find here.}

Iterative minimization found asymmetric conic states along the phase boundary between
the (1/2,1/2,1/2) antiferromagnet and the (1/2,0,0) antiferromagnet, but we
believe these are artifacts, in the sense we will describe.
This boundary corresponds to LT modes degenerate over a plane of wavevectors,
leading to a degenerate family of spin ground states with an arbitary wavevector. 
These are generically non-coplanar spirals,
except the limiting states of this family are collinear ``encompassed states'' 
(in the nomenclature from Sec.~\ref{sec:bridging}).
Thus, although these conic spirals are valid ground states, we do not count
them as non-coplanar, since that is not {\it forced} by the couplings.
This is an instance where the overlap between ``encompassing'' and ``degenerate'' states is especially stark, as the family of degenerate states coincides with the class of encompassing states.

The impossibility of forcing {\it any} conic spiral
in the octahedral lattice is understood by using the 
mapping ~\eqr{eq:mapping} of couplings from the 
octahedral lattice to the chain lattice (see Sec.~\ref{sec:mapping}).
We will see shortly (Sec.~\ref{sec:chain})
that stabilizing either kind of conic requires a coupling $j_3$ or $j_4$ 
in the chain lattice; for a (100) stacking vector,
Eq.~\eqr{eq:mapping} takes octahedral coupings $J_1$ through $J_4$ 
to chain-lattice couplings $j_0$ through $j_2$, so
clearly couplings $J_5$, $J_6$ or longer  are required (and sufficient)
to truly stabilize conic spirals in the octahedral lattice.

\SAVE{It was near the triple point of the $\pi$/3 cuboctahedral state,
(1/2,1/2,1/2) antiferromagnet, and (1/2,0,0) antiferromagnet,
that we observed the double-twist state.
The couplings originally used to produce this state were $J_1=-2$, $J_2=-3.771657$, $J_3$=1 
(all other couplings 0). 
Given its proximity to a triple point and the small size of the region 
where $(qq0)$ optimizes the $\kkLT$, it is likely that the placement
of the double-twist region in the phase diagram 
is topologically analogous to the helimagnetic region. 
That is, there is a small finite region (near a triple point) 
where it is stable.}

\subsection{Chain Lattice}
\label{sec:chain}

The chain lattice ground states are significantly more complicated than those of the octahedral lattice. Analytically determining the optimal energy of even the helimagnet becomes difficult when couplings beyond $j_3$ are included. Therefore, while we can easily determine a variational form for the energies, we cannot analytically determine the ground state when couplings $j_3$ or higher are introduced. Energies are given in table \ref{tab:C-States}, which are then numerically optimized to give the subsequent phase diagrams. We once again normalize by $|j_1|$, but we now plot $j_2'/|j_1|\times j_2/|j_1|$, rather than $J_2/|J_1|\times J_2'/|J_1|$. This change of convention does not have great physical implication, as the difference between $J_2$ and $J_2'$ in the octahedral lattice is distinct from the difference between $j_2$ and $j_2'$. Lastly, by the definitions of the chain lattice, several properties of the phase diagram follow immediately. First, simultaneous exchange of $j_2$ with $j_2'$ and $j_4$ with $j_4'$ will merely change the labeling convention to distinguish the two sublattices. The ground state in the chain lattice must therefore be invariant under this operation. Furthermore, when $j_2=j_2'$ and $j_4=j_4'$, this exchange will not change anything. In this region of parameter space, there is no difference between the two sublattices and the chain lattice becomes a Bravais lattice with unit cell 1/2. 
From this fact and Sec.~\ref{sec:LT}, it follows that states in this region are necessarily coplanar.

\begin{table}
\begin{tabular}{|c|c|c|}
\hline 
State & $Q$ &  Energy\tabularnewline
\hline
\hline 
Ferro & 0 & $ -2j_1-j_2-j_2'-2j_3-j_4-j_4'$  \\
\hline 
AFM & 0 & $2j_1-j_2-j_2'+2j_3-j_4-j_4'$ \\
\hline 
Heli- &  2$\psi$ & $-2j_1\cos\psi-(j_2+j_2')\cos 2\psi$ \\
magnet &  &  $-2j_3\cos 3\psi-(j_4+j_4')\cos 4\psi$ \\
\hline 
splayed & 0 and & $(j_1+j_3)^2/2j_2+j_2-j_2'-j_4-j_4'$ \\
ferro   & 1/2 & \\
\hline 
splayed & 0 and &  $ (j_1+j_3)^2/2j_2+j_2-j_2'-j_4-j_4'$ \\
ferri   & 1/2 & \\
\hline 
Alter- & 1/2  & $-2(j_1\cos\psi+j_3\cos 3\psi)\cos\alpha$\\
nating & and & $+ j_2-j_4-j_2'\cos 2\psi-j_4'\cos 4\psi$  \\
Conic  & 2$\psi$ & $-(j_2[\cos 2\psi+1]+j_4[\cos 4\psi-1])\cos^2\alpha$ \\
\hline 
Asym- & 0 & $-2(j_1\cos\psi+j_3\cos 3\psi)\cos\alpha \cos\beta$ \\
metric & and  &  $+  2(j_1+j_3)\sin\alpha \sin\beta-j_2-j_2'-j_4-j_4'$ \\
conic & 2$\psi$   &  $-(j_2\cos^2\alpha+j_2'\cos^2\beta)(\cos 2\psi-1)$\\
    &   &   $ -(j_4\cos^2\alpha+j_4'\cos^2\beta)(\cos 4\psi-1)$\\
\hline
\end{tabular}
\par
\caption{Parametrizations and Energies of the Chain Lattice.
$Q$ denotes the ordering wavevector(s), as a multiple of $2\pi$.
The energy per unit cell is given.
The splaying angle is given by $\cos\alpha  = -(j_1+j_3)/2j_2$ for
the splayed ferromagnet or $+(j_1+j_3)/2j_2$ for the splayed ferrimagnet.
}
\label{tab:C-States}
\end{table}

To classify phase boundaries in the chain lattice, it is important to consider limiting cases 
(i.e. encompassed states in the nomenclature of Sec.~\ref{sec:bridging}). 
States with variational parameters (the helimagnetic wave-vector $q$, as well as the 
conic/splay angles) will undergo a second-order transition when their parameters reach a 
limiting value (0 or 1/2 for the helimagnet angle, 0 or 1/4 for conic angles). 
Thus there is a second-order transition between helimagnetism and either 
antiferromagnetism or ferromagnetism, as well as between alternating conics 
and every other state (except asymmetric conic). 
Asymmetric conics, on the other hand, have second-order transitions to ferromagnetism, 
antiferromagnetism, helimagnetism, or ferrimagnetic splayed states, 
but not to ferromagnetic splayed states or alternating conics. 
The splayed states, meanwhile, can only have second-order transitions to 
ferromagnetism or antiferromagnetism (depending upon which type of splayed state it is), 
to the other splayed state, or the appropriate conics. 
All other transitions are necessarily first-order.

\SAVE{The splayed phases were established through numerically optimzation. 
They had first been observed through iterative minimization, 
but dismissed as obviously artifacts.}

\begin{figure}
\includegraphics[scale=0.6]{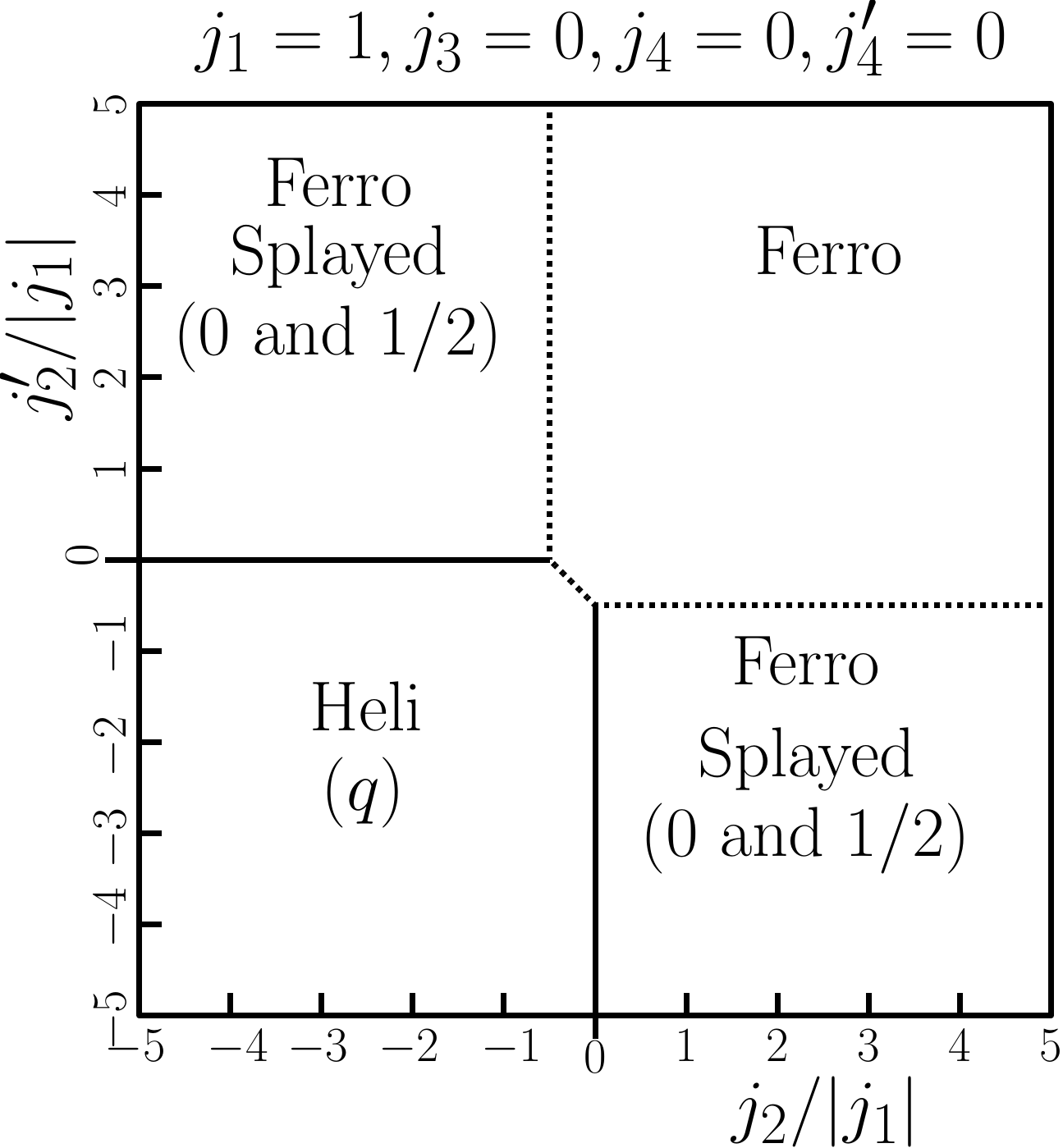}
\caption
{Chain Lattice Phase Diagram for $j_1$=1, $j_3$=0,
$j_4$=0, $j_4'$=0. Note similarity of phase boundaries
to Figure \ref{fig:O,1,0}, despite the lack of other similarity.}
\label{fig:C,1,0,0}
\end{figure}

Consider first the case of $j_1$ ferromagnetic with no couplings beyond $j_2,j_2'$ (Figure \ref{fig:C,1,0,0}). The phase diagram displays the same quadrant structure that we found in the octahedral lattice. However, the quadrant structure is not identical in the two lattices. First of all, the ground states are different in the chain lattice (helimagnetism and splayed states instead of various forms of antiferromagnetism). Secondly, the topology of the first and second order transitions are reversed for the two lattices. Both of these phenomena can be explained by appealing to the additional degrees of freedom in the octahedral lattice. Because the octahedral lattice has three spatial variables, it has ground states that cannot exist in the chain lattice. This includes families of degenerate states at the phase boundaries of the octahedral lattice, producing second order transitions (when there are second order transitions in the chain lattice, they are principally due to encompassing states).

\SAVE{The older versions of Figures \ref{fig:C,-1,28,0}
and \ref{fig:C,-1,0,-28} included a 
dash-dotted line indicating ``range of 
values used in iterative minimization''
(after dividing by $|j_1|$). In
Figure~\ref{fig:C,-1,28,0} this has slope 1/2 and
is within the Heli phase. It is marked
``observed numerically by varying $j_3$''.
In Figure \ref{fig:C,-1,0,-28}, there are 
two such lines having slopes $\pm 1.5$,
forming an X shape in the Heli phase but
just crossing into both the adjacent conics.
These are marked ``observed by varying $j_4$''.
CLH's interpretation: these phases were
reached by taking $j_1\approx j_2 \approx j_2' \approx -1$
and varying $j_3$ or $j_4$, which we understood to be
essential for getting a non-planar result.
But does that include
the asymmetric conic in Figure~\ref{fig:C,-1,28,0}?
The dot-dashed line, in that figure, was not shown
crossing into the asymmetric conic.
SS explains: we did not observe the asymmetric conic in
Figure ~\ref{fig:C,-1,28,0} through iterative minimization
(nor were we likely to find it by picking random numbers, since
it's a relatively narrow slice). Those conics were found through
numerically optimization the variational forms.}

\begin{figure}
\includegraphics[scale=0.6]{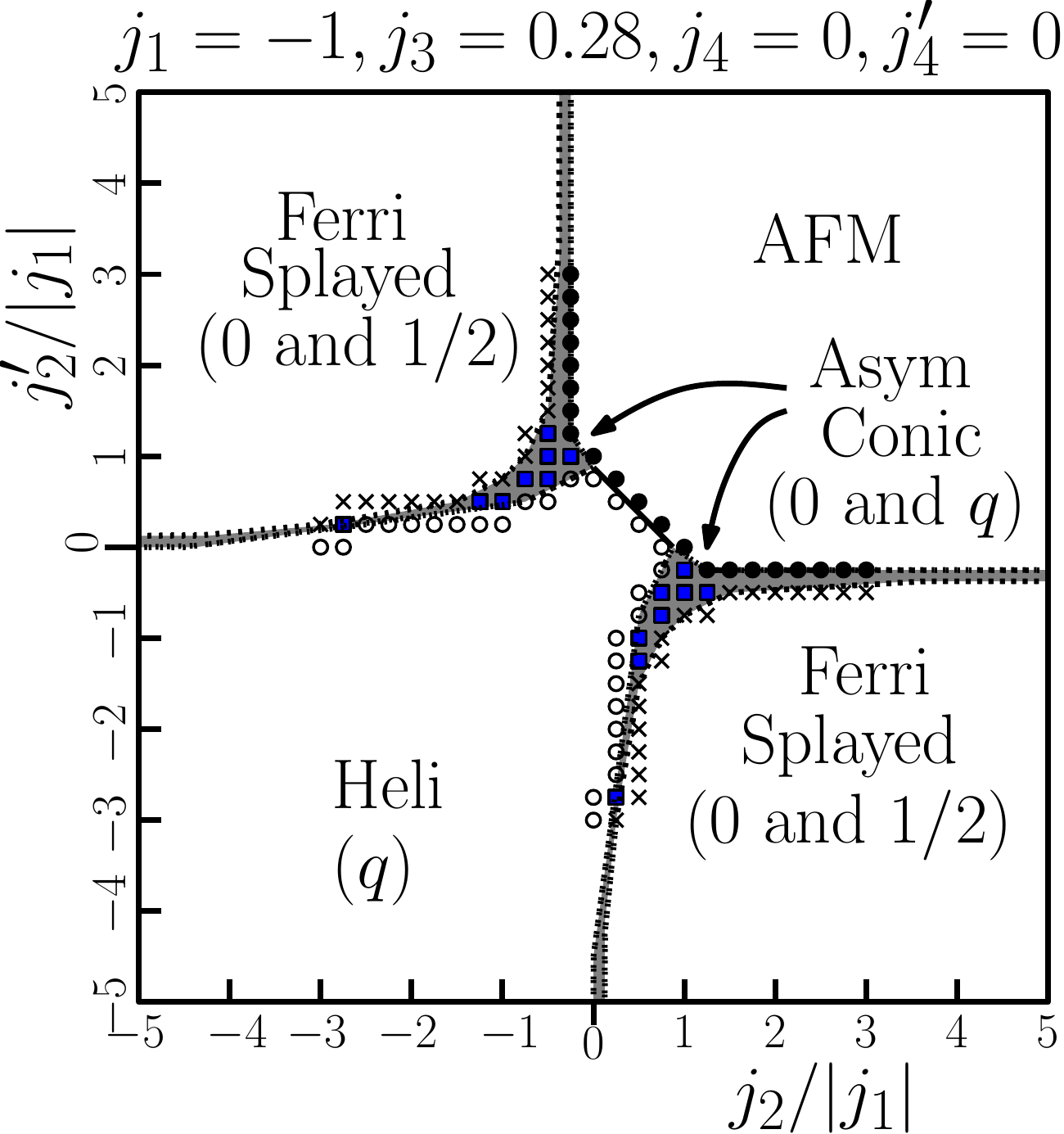}
\caption
{(COLOR ONLINE) Chain Lattice Phase Diagram for $j_1=-1$,
$j_3$=0.28, $j_4$=0, $j_4'$=0. First state with asymmetric
conic, and therefore first instance of non-coplanar ground states
in the chain lattice. 
White circles denote helimagnetic phase,
black circles anti-ferromagnetic,
crosses splayed ferrimagnetic, and 
filled squares (blue online) asymmetric conic,
the phases determined by numerically optimizing the variational form
of the energy for these couplings.
}
\label{fig:C,-1,28,0}
\end{figure}

Next, we examine the case $j_1$ antiferromagnetic and $j_3$ ferromagnetic (Figure \ref{fig:C,-1,28,0}). Several states in the quandrant structure are different from the ferromagnetic $j_1$ case (i.e. ferrimagnetic vs ferromagnetic splayed), but more interesting is the presence of the asymmetric conic - the first instance of a nonplanar state in the chain lattice. In much the same way that tuning $j_3$ in the octahedral lattice produced helimagnetic states around the phase boundaries of the more common states, so in the chain lattice do we find that the asymmetric conic state becomes stabilized around what would be the antiferromagnetic, ferrimagnetic splayed, helimagnetic triple point. And unlike the helimagnetic state in the octahedral lattice, which had both first order and second order transitions, the asymmetric conic has only second order transitions (this is because it is an encompassing parametrization of every other state in this slice of the phase diagram).


If we switch the sign of $j_3$ so that we have 
both $j_1=-1$ and $j_3=-0.28$ antiferromagnetic, the topology
of the phase diagram (not shown) is much the same as Figure~\ref{fig:C,-1,28,0}.
The AFM/Heli boundary gets shorter and moves
towards the lower left, so that the four domains almost meet at a point.
More significantly, the asymmetric conic does not appear along the phase
boundaries. Instead, all transitions are continuous, except that along both 
parts of FerriSplayed/Heli boundary, the portion closest to the center
is first-order.~\cite{FN-spurious-AC}
(The point where the nature of the transition
switches from first-order to continuous is thus of tricritical type.) 

\SAVE{This artifact was still in the thesis/paper as of summer 2010.
Matt Lapa's early learning experiences with the chain lattice brought
out the discrepancy, resolved in some emails roughly Nov. 2010.
SS elaborated 8/30/12:
(i) we had a range of numerically optimized $j_i$ values for which a conic was the
predicted ground state.
(ii) SS selected a set of $j_i$'s within this domain for which the math was simpler
(i.e. rational numbers) and plugged them into variational form of the conic
energy. Recall the conic ``encompasses'' the planar states which are its limiting
cases, and which are the competing states.
So if we properly optimize the conic, it will definitively specify the 
ground state for those parameters.
(iii) SS analytically optimized the variational form of the conic energy. 
Result: with the analytically determined optimal $\alpha,\beta,\psi$, the conic 
states are slightly higher in energy (not by much)
than the planar helimagnetic states, which are the true ground states.}

\SAVE{Elaboration: this problem is a competition between helimagnetic and conic phases. 
Initially, to find the phase boundaries, we sampled the $\psi$ values
within a grid, and pick whichever was numerically found as the minimum.
Given a trial value of $\psi$, for MOST values of $\psi$, a conic state 
is the minimum.  However, there is a very small interval
where it crosses over to helimagnetic, just barely -- and
(ADDED) that interval includes the absolute optimal $\psi$.
The difference in energy is really tiny there, since both phases reach their minima at 
nearly the same Q-point. Because this range is so small,
a fixed grid is unlikely to hit upon this region, and we
would incorrectly identify the optimum state as being the conic.
The correct result was found after analytically optimizing.
Selecting coupling values from within this pocket and analytically optimizing
the variational form of the conic shows that the ground state within these
pockets should actually be splayed.
(CLH *thinks* ``analytically optimizing'' means solving 
$\partial_\alpha E = \partial_\beta E = \partial_\psi E=0$
but is not quite sure about this.) (SS: CLH is correct.)}

\begin{figure}
\includegraphics[scale=0.6]{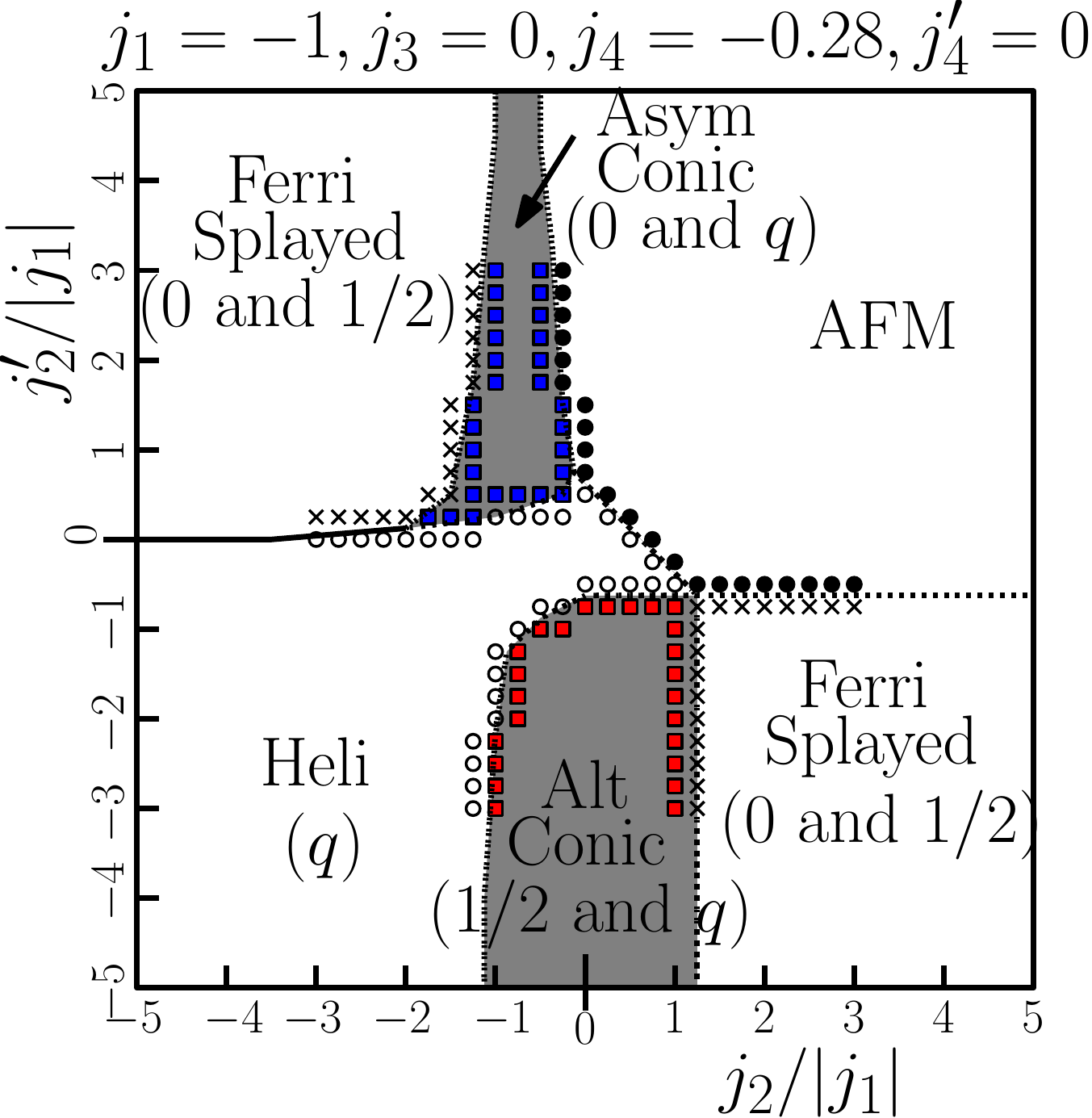}
\caption
{(COLOR ONLINE) Chain Lattice Phase Diagram for $j_1=-1$,
$j_3$=0, $j_4=-0.28$, $j_4'$=0. 
Filled squares denote conic spiral phases, of which 
both kinds exist in this parameter slice:
symmetric conic in the upper half (blue online)
and asymmetric conic in the lower half (red online).
This is our first instance of an alternating conic.
\SAVE{(It shares precedence as the first instance of
a non-coplanar state in the chain lattice with the asymmetric conic
state in Figure \ref{fig:C,-1,28,0}).}
}
\label{fig:C,-1,0,-28}
\end{figure}

Finally, we consider the case of $j_1$ ferromagnetic and $j_4$ antiferromagnetic (Figure \ref{fig:C,-1,0,-28}). This slice of parameter space is quite interesting as both types of conics are present. Furthermore, the alternating conic now fills a relatively large region of parameter space. This is likely due to its highly non-linear dependence on $j_4$ (as a function of its variational parameters).

\section{Conclusion and discussion}
\label{sec:discussion}

To conclude the paper, we first review our principal results, 
and then assess how much of what we learned is transferable to
other lattices.

\subsection{Summary}

The highlights of this paper include both concepts and methods, 
as well as results specific to the octahedral lattice, which
seems relatively amenable to non-coplanar states.
We pay special attention (Sec.~\ref{sec:scarce}) to commonalities 
in the positioning of non-coplanar states in the phase diagram
vis-\`a-vis neighboring phases.

Our overall focus had a flavor of reverse engineering, in that we 
try to ask which couplings gave a certain phase (or which gave 
any non-coplanar phase) -- of course in order to do that, one must
also understand the forwards question (given the couplings,
what is the phase).
In that sense, our work is an example of a ``materials by design'' 
philosophy, whereby materials are tailored -- e.g. by adjusting their chemical
content --  to have a combination of interactions leading to
a desired state.

\subsubsection{Methods}

Our basic recipe to determine the ground state of a non-Bravais lattice 
was a two-step process (Sec.~\ref{sec:framework}). 
First, an approximate ground state configuration
is generated through iterative optimization (Sec.~\ref{sec:iter-min}) 
of a lattice, starting from a random initial spin configuration. 
From this result, an idealized spin configuration is created. 
The idealized formulation, if it has parameters 
undetermined by symmetry, is then used variationally optimize 
the Heisenberg Hamiltonian (Sec.~\ref{sec:var}), yielding
also the energy per site as a function of parameters.
When this has been carried out for each candidate phase, 
a phase diagram (Sec.~\ref{sec:phase}) can be generated.

We refined the basic recipe 
further using three additional concepts or tricks.
First, although the eigenmodes of the coupling
matrix do not (in the non-Bravais case) automatically
lead us to the ground  state, they sometimes do work
and are always a useful guide.
Second, if the magnetic structure is layered, 
the three-dimensional octahedral lattice can be 
mapped to various (non Bravais) one-dimensional lattices; 
such  ``chain'' lattices are much more tractable than
the cuboctahedral one, but they still support non-coplanar states.
Third, we applied perturbation analysis to find second-order 
phase boundaries
(especially in combination with variational optimization,
but also in the Luttinger-Tisza Fourier analysis).
Namely, when a ground state could not be written exactly,  but 
emerges as an instability of a closely related state
-- e.g. a helimagnetic phase out of the ferromagnetic one --
we can expand around the latter state and solve for the couplings
at which it goes unstable.

The insufficiency of finding regions with the correct $\kkLT$ 
for the purpose of stabilizing a targeted magnetic structures 
is an important caveat for the LT approach. There are some combinations
of modes which {\it cannot} be stabilized by {\it any} set of pairwise interactions.
That is, we can make them be optimal modes, but only if
they are part of a degenerate family that includes other
optimal LT modes that are unrelated by symmetry.

\subsubsection{Results}

We identified two categories of non-coplanar arrangements
(which could be defined for more general lattices than the
octahedral one).
First, there are the commensurate three-$\QQ$ states,
exemplified here by the highly symmetric 
``cuboctabedral'' spin state (Sec.~\ref{sec:cuboc}).
A second general class includes several varieties of
incommensurate  ``conic spiral'' (Sec.~\ref{sec:conic}).  
The commensurate three-$\QQ$ states were found for both short and long range 
couplings (i.e. $J_i$'s limited to no more than second-nearest neighbors 
or extending beyond second-nearest neighbors), while incommensurate conic spirals
were possible for our lattice only when there 
are some $J_i$'s beyond the second neighbors.

The conic spirals came in ``alternating'' or ``asymmetric''
subclasses (stable in different parts of $\{ J_i \}$ 
parameter space) according to whether the spin components 
along the rotation axis were the same in each layer
or alternated.  
This suggested that an additional subclass should exist, 
the ``transversely modulated conic spiral'',
which is non-uniform in each layer.
In addition, we came across a ``double-twist'' ground state
of high (but not cubic) symmetry, which has some commonalities
with the cuboctahedral state but is probably best classified
as a transverse-modulated conic spiral
(Sec.~\ref{sec:doubletwist}).

\SAVE{Performing this analysis gives two classes of non-coplanar ground states. The first is the commensurate triple-Q state, which includes configurations like the cuboctahedral states (Sec.~\ref{sec:cuboc}). These states mix different LT modes that are all related by symmetry, giving a highly symmetric ground state (but these symmetries are not always very obvious, as in the double-twist state (Sec.~\ref{sec:doubletwist}), which bears some resemblance to the triple-Q states).}

\SAVE{The second class is the incommensurate conic spiral states (Sec.~\ref{sec:conic}), where the ground state is a linear combination of two modes: an arbitrary helimagnetic (normally with $\vec{q}$ along (100) or a permutation) and a ferromagnetic or antiferromagnetic state (i.e. states with wave-vectors confined to the boundaries of the Brillouin zone).}

The conic spiral states were found to be a function of one spatial coordinate, allowing the more involved octahedral lattice to be mapped to the 1D chain lattice while preserving the ground state (Sec.~\ref{sec:mapping}). While this mapping is not a generic property of ground states in the octahedral lattice, it dramatically simplifies the analysis. Moreover, the transformation provides a guide for the couplings required to stabilize the conic spiral states in the octahedral lattice.

Note that while the presence of inequivalent sublattices was necessary for the stabilization
of non-coplanar states in the one-dimensional chain lattice, it is not so in higher dimensions.
Our octahedral-lattice couplings explicitly treat all sites equivalently; non-coplanar states
emerge either when it divides into unequal layers, that map to the chain lattice,
or when it supports three-dimensional spin patterns not mappable to a chain lattice.

The most dramatic examples when the chain lattice fails to represent 
the ground state in the octahedral lattice are the cuboctahedral states,
for which there is no distinguished direction of variation (or stacking vector) 

\subsubsection{Why noncoplanar states are scarce}
\label{sec:scarce}

One might have expected noncoplanar ground states 
to be generic (in the non-Bravais case, when they are possible at all)
but in fact they were seen in only a 
small portion of our parameter space (Sec.~\ref{sec:phase}).
In fact, non-coplanar phases typically appeared in the phase diagram 
as ``bridges'' intermediate between simpler phases; there are two 
scenarios of bridging, as formulated in Sec.~\ref{sec:bridging}.

First, typically along the phase boundary between two 
\SAVE{or more} simple commensurate phases, one finds
families of continuously degenerate states.
Usually, this infinite family includes coplanar states
but the generic member is non-coplanar; appropriate
tuning of the couplings (e.g. including further neighbors) 
can select particular non-coplanar states in different ways. 
However, since the non-coplanar phase was limited
(in the first-order description) to the (measure zero)
boundaries between planar states, it naturally occupies only
in a small region of the extended parameter space. 
Indeed, the non-coplanar states we found in the octahedral lattice
either occurred in small wedges, e.g. the cuboctahedral or double-twist states 
or were only observed as accidental instances of the degenerate family
in cases where the degeneracy cannot be broken (within the parameter 
space we took as the scope of this paper), 
e.g. the asymmetric conic states.

Second, there are the ``encompassing states''. Such states become qualitatively
different as a free parameter reaches some limiting value 
(a canonical example is the collinear ferromagnet 
as a limiting case of planar spirals.)
In the chain lattice, we often observed phase boundaries 
between two states that did not encompass each other; by tuning the couplings, 
it was sometimes possible to stabilize a third state 
which encompassed the original two, so the original first-order transition 
is converted to two successive second-order transitions with the 
encompassing state in the middle.  This is the means by which 
conic spiral states are stabilized in the chain lattice.

Recently, a framework was proposed with a motivation similar to ours,
the ``regular states''~\cite{messio-lhuillier} which have a magnetic
symmetry such that all sites symmetry-equivalent.  The cuboctahedral
state is an example.  However, this is neither necessary nor sufficient
for our own problem, to find all kinds of non-coplanar states:
most of those, e.g. the double-twist state, are not regular, and
conversely certain non-coplanar regular states are
not stabilized by bilinear exchange interactions alone.

\subsection{Generalizations to other lattices?}

Now we consider moving beyond the somewhat artificial octahedral lattice.
We examine the necessary conditions for non-coplanar spin configurations
analogous to the two main classes we discovered (cuboctahedral and conic
spiral).  Can our results be applied to other lattices,
such as the pyrochlore lattice?

\subsubsection{Generalizing the cuboctahedral state?}
\label{sec:generalize-cuboc}

The cuboctahedral states seem highly specific to the octahedral lattice,
since they possess the same symmetries as the lattice.
On the other hand, the LT construction (each sublattice using different combinations of LT modes, the combinations being related by rotational symmetry of the different sublattices) seems fairly generic.
A general name for states like the cuboctahedral state
might be ``commensurate triple-Q state'', referring to
its content of LT modes.
Can such states be found in other lattices, or are there others in the
octahedral lattice?

One answer requires that the hypothetical generalized state 
enjoys the full symmetry of the lattice,
as the cuboctahedral state does.
Then the number of site classes must be a multiple of the number
of spin components, i.e. of three.
That will not work for
the pyrochlore lattice, in which there are {\it four} site
classes associated with (111) directions.  However, it
might work in the half-garnet lattice, which has
{\it six} site classes associated with (110) directions.

A second, more systematic way to answer question follows 
the LT approach of Sec.~\ref{sec:cuboc-LT}.  The state
must be a linear combination of three modes -- one for
each component of spin.  For the state to enjoy the
full lattice symmetry, these must be the complete star
of symmetry-related modes, thus it must have a 
threefold multiplicity.  Furthermore, each mode must be
real (otherwise one mode requires two spin components);
that happens only if the wavevector is half a reciprocal
lattice vector, i.e. is at the center of one of the 
Brillouin zone's faces.   

However, Sec.~\ref{sec:cuboc-no} gave a cautionary example:
finding such modes is not {\it sufficient}, because it 
might give a state with decoupled sublattices, so that the
cuboctahedral state would merely be one undistinguished
configuration in a continuous manifold of degenerate states
that even includes collinear states.  We can, in fact,
borrow a notion from Section~\ref{sec:mapping} to guess when
this happens: mapping to a chain lattice, now applied to LT
modes rather than to spin structures.  The wavevector of one 
mode defines a set of planes, and thus a way of projecting 
both the sites and the mode onto a one-dimensional chain lattice.   
On the chain lattice, in order that the mode be real, 
the wavevector must be $\pi$.  If all the site planes were equivalent, 
the chain lattice is a Bravais lattice and the mode must be a plane
wave; with $Q=\pi$, it is easy to see we get a decoupled pattern,
namely $(+1,0,-1,0,...)$.  In the octahedral lattice,
the $\{1/2,0,0\}$ modes have a threefold degeneracy and also
correspond to a non-equivalent set of stacked planes, giving
a cuboctahedral state, whereas the 
threefold degenerate $\{1/2,0,0\}$ modes have equivalent
stacked planes and give decoupled states.
In the pyrochlore lattice, the $\{1,0,0\}$ modes have a threefold
multiplicity but have equivalent stacked planes; on the other hand,
the $\{1/2,1/2,1/2\}$ modes have inequivalent stacked planes
but their multiplicity is fourfold.  Thus, in the pyrochlore
case this path does not lead us to a cuboctahedral state.

The third way to answer this question is via the cluster constuction of
Sec.~\ref{sec:cuboc-cages}, in which the lattice was
decomposed into cuboctahedral cages.
Indeed, some other lattices consist of a union of
roughly spherical ``cage'' clusters:
e.g., in the Cr$_3$Si-type or ``A15'' structure, the majority 
atoms form cages in the form of distorted icosahedra.
Furthermore, couplings can certainly be chosen such that
the spin ground state of a cluster forms a the same
polyhedron as the cluster itself.

However, our ground-state construction (Sec.~\ref{sec:cuboc-cages})
demanded that the spin configurations in
adjacent cages be related by a {\it reflection}
in spin space. Thus the cages alternate between proper
and improper rotations of a reference configuration.
That is possible only if the cage centers form a 
{\it bipartite} lattice, which is not true for the 
A15 majority-atom cage centers.  (They
form a bcc lattice with first- and second-neighbor links.)

\SAVE{And it does not seem likely, when you consider that
``adjacent'' should be defined in a Voronoi sense: that
generically produces triangles of nearest neighbor bonds.}

\subsubsection{Generalizing the non-coplanar spirals?}
\label{sec:disc-generalize-spirals}

Long ago, Kaplan and collaborators studied 
{\it doubly} magnetic spinels, and identified a
(noncoplanar) ``ferrimagnetic spiral'' configuration
arising from exchange interactions (a spiral of
this sort is responsible for multiferroic
properties of CoCr$_2$O$_4$~\cite{kaplan-CoCr2O4}.
This is actually a kind of double spiral
(each magnetic species accounting for one of the
spirals).  Notice that we could map that structure
to a chain, too, and that chain would have two 
inequivalent sites, which we point out is a 
precondition for developing a noncoplanar state.

For a mapping to the 1D chain lattice to provide a non-coplanar state, 
the mapped sites must be inequivalent by translation;
if not, the chain lattice is a Bravais lattice and must
have (at most) a coplanar spiral state.
This in turn depends on having {\it unequal} layers in the three-dimensional lattice,
which is is impossible in the Bravais case,
but inequivalent layers can emerge from a non-Bravais 
lattice with fully symmetric couplings.
Thus, as we worked out in Sec.~\ref{sec:mapping},
a $\{100\}$ stacking in the octahedral lattice 
has twice as many sites on $x$ and $y$ bonds, constituting one
kind of layer, as there are sites on $z$ bonds,
constituting the other kind.

The stacking direction is one in which the wavevector is
incommensurate, and so we may seek out parameter sets
for which $\kkLT$ goes incommensurate in a desired direction.
However, this is no guarantee that the actual ground state
is stacked, since it might be better energetically to combine
these modes (or ones nearby in $\kk$-space) in a quite different way. 
Thus, although we found some optimal $\kkLT$ wavevectors in
directions other than $\{100\}$, they were never the basis
of a stacked spiral.

\subsection{On to pyrochlore lattice?}

The pyrochlore lattice is undoubtedly the most prominent
non-Bravais lattice that is likely to have non-coplanar
states, and was a major motivation for the methods we developed
in this paper for the more tractable octahedral lattice.

Relatively few works have tackled the isotropic spin problem on
the pyrochlore lattice, beyond the (massively degenerate) case of only $J_1$
[analogous to the $J_1=J_2$ case on the cuboctahedral lattice].
The $J_1$--$J_2$, or $J_1$--$J_3$ pyrochlore, with large antiferromagnetic
$J_1$, has a noncoplanar and somewhat obscure state that is not 
fully understood at $T=0$~\cite{Tsu07,Na07,Che08}.
In the spinel GeNi$_2$O$_4$, neutron diffraction found 
a $\{111 \}$ ordering consisting of alternating kagom\'e
and triangular lattice layers with different densities of spins,
reminiscent analogous of the $\{100\}$ stackings we have
explored in the octahedral lattice.
Ref.~\onlinecite{Ma08} invoked interactions out to $J_4$ 
in order to rationalize this state, but did not verify it was 
the ground state for the suggested interactions.
Finally, to address metallic pyrochlore compounds, a toy Hamiltonian 
was investigated on the pyrochlore lattice with exchange couplings
having RKKY oscillations~\cite{kawa-RKKY}.

Very recently, a comprehensive study has been carried out for just
the $J_1$--$J_2$ phase diagram of the pyrochlore lattice by applying
the methods of the present paper, which were equally effective
in that case~\cite{lapa}.  About four complex states were identified,
including that above mentioned state introduced by Kawamura and 
collaborators~\cite{Tsu07,Na07,Che08}; a state reminiscent
of our double-twist state; 
a more complex generalization of a conic spiral;
and a cuboctahedral spin state that
has less than cubic lattice symmetry, 
thus evading the negative conclusions of Sec.~\ref{sec:generalize-cuboc}
\cite{lapa}.

\begin{acknowledgments}

We thank G.-W. Chern, J-C. Domenge, C. Fennie, S. Isakov, 
T. Kaplan, M. Lapa, and M. Mostovoy
for discussions and communications.
C.L.H. was supported by NSF grant DMR-1005466; S. S. was supported by
the Cornell Engineering Learning Initiatives Program, the Intel Corporation, 
and the Rawlings Cornell Presidential Research Scholars Program.

\end{acknowledgments}

\appendix
\section{Perturbative Calculation of Phase Boundaries}
\label{app:phase-boundaries}

To move from a collection of ground states discovered at discrete points in parameter 
space to draw a full phase diagram is non-trivial.
To analyze the phase boundaries of the more complicated states (those with free parameters in their parametrization)
 we depend on either variational optimization
(Sec.~ref{sec:var}), or some kind of perturbation theory.
Perturbation theory can be applied in two places: either to the LT matrix,
or to the Hamiltonian of a parametrized spin state. 
The former is more straightforward, but is limited since most of the
non-coplanar ground states are not exactly built from optimal LT 
eigenmodes.

An obvious caveat for either of these applications of perturbation is that they 
detect continuous transitions, representing infinitesimal changes in the 
spins: it is a bifurcation of the local minima as points in the 
configuration space.   But there the ground state might instead change 
due to a first order transition, when the energies of two 
separated configurations cross as parameters are varied.
We do encounter the first-order case on occasion, though not nearly so 
often as the continuous one.
To detect such discontinuous transitions, we must compare numerical 
calculations of the ground state energies.

Our original question was ``given a certain set of parameters,
what is the ground state'', but in these calculations it has
been turned around to ``given a particular ground state, 
for what parameter sets is it favored?''

\subsection{LT Analysis (octahedral lattice example)}
\label{sec:pert-LT}

We give an example here of the use of 
perturbation theory to discover the incipient instabilities
of LT modes.  Such an approach may be especially useful to 
locate the phase boundaries for transitions from an LT 
exact ground state to more complicated, incommensurate states.  
If one is hunting for the parameter domain which would stabilize
an particular mode with ordering wavevector $\QQ$, a substitute
problem is to find the parameter domain in which this mode
(or one with $\kkLT$ similar to $\QQ$) is the optimal LT mode.
We reiterate the caveat from Sec.~\ref{sec:LT}:
such a discovery  is a necessary but not sufficient criterion 
to guarantee the existence of any ground state based on the obtained $\kkLT$.
(It is not sufficient because the actual ground state could feature additional modes or modes merely in the neighborhood of $\kkLT$.)

\SAVE{This corresponds best to the actual 
ground states of the octahedral lattice then they are planar spirals
of some form, characterized by a single eigenmode with a unique $\QQ_0$.}

The LT matrix elements $J_{ij}(\kk)$ and eigenvalues  $\tJ(\kvec \nu)$
are functions of wavevector $\kk$.
Imagine that $\QQ_0$ is a point of high symmetry in the zone so as to
be a stationary point for $\tJ(\kvec \nu)$.  For some parameter sets,
we know, it is a minimum and in fact optimal; whereas for some other parameter
sets, we imagine, it is only a saddle point, and the minimum occurs at
some nearby wavevector of lower symmetry.

One first writes a Taylor expansion of the LT 
matrix in powers of $\delta\kk\equiv\kk-\QQ_0$.
Using standard techniques (formally identical to those used
for eigenfunctions of the Schr\"odinger equation)
it is straightforward to write a perturbation expansion for
$\tJ(\kvec \nu)$ in powers of $\delta \kk$.
Inspection then shows where this stops being positive definite.
Since the LT matrix elements are bounded, 
so are the eigenvalues $\tJ(\kvec \nu)$.
So if the mode at $\QQ_0$ goes unstable
at quadratic order in $\delta \kk$, 
there must be higher-order positive terms in the expansion.
Thus the single local minimum of $\tJ(\kvec \nu)$ bifurcates in some
fashion.  The corresponding spin state cannot be a commensurate spiral,
but it might be representable in the framework of planar stackings
(Sec.~\ref{sec:mapping}).

This technique was used, for example,  to analyze how the 
$\QQ_0=(1/2,1/2,0)$ wavevector is destabilized in the LT phase diagram.
Simulations had found degenerate antiferromagnetic orderings
at that wavevector; if an incommensurate wavevector of form $(q,q,0)$
had been stabilized, this might have been the basis of a non-coplanar
spiral stacked in the $(1,1,0)$ direction.

The LT matrix is given by:
    \begin{subequations}
       \begin{align}
        J_{ii}(\kk) &=&
          2 J_2 \cos k_i + 2 J_2' (\cos k_j + \cos k_k) + 4 J_4 \cos k_i  \nonumber \\
      &&  \times (\cos k_j + \cos k_k) + 4 J_4' \cos k_j \cos k_k
     \label{eq:LT-matrix-diag}  \\
          J_{ij}(\kk) &=&
            4 J_1 \cos \frac{k_i}{2} \cos \frac{k_j}{2}
     + 8 J_3 \cos \frac{k_i}{2} \cos \frac{k_j}{2}  \cos k_k.
     \label{eq:LT-matrix-offdiag}  
           \end{align}
     \label{eq:LT-matrix}
     \end{subequations}
Here $j,k$ in \eqr{eq:LT-matrix-diag} means the indices other than $i$,
similarly $k$ in \eqr{eq:LT-matrix-offdiag} is the index other than $i,j$.
If we substitute $\kk=(k,0,0)$, for example, the eigenvalues along this cut are
    \begin{subequations}
\begin{widetext}
 \begin{align}
\tJ(\kvec, 0)  = & 
      -J_2+2J_4'-(J_F(0)-J_2)[1+\cos k]+2J_X(k); \\
\tJ(\kvec, \pm) = & 
         -\frac{1}{2}  \Big[2J_2'-4J_4+(J_F(0)+4J_4)(1+\cos k)+2J_X(k) 
  \Big] \nonumber \\
    & \pm\frac{1}{2} \Bigg\{ 
        [(J_2-J_2'+2J_4-2J_4') (1-\cos k) 
                      +2J_X(k)]^2+
        16J_X^2(0)(1+\cos k) \Bigg\} ^{1/2}.
   \end{align}
\end{widetext}
   \end{subequations}

Along another cut through the zone, $\kk = (k,k,0)/2\sqrt{2}$, the eigenvalues are
    \begin{subequations}
\begin{widetext}
 \begin{align}
\tJ(\kvec, 0)  = & 
      J_2-2J_4'-2J_F(0)\cos^2 k+2J_X(0)\cos^2 k; \\
\tJ(\kvec, \pm)  = & 
         J_2'+2J_4\cos^2 2k-(J_F(2k)+2J_2'+4J_4\cos 2k)\cos^2 k-2J_X(0)\cos^2 k 
 \nonumber \\
    & \pm\frac{1}{2} \Bigg\{ 
        [(J_F(2k)-4J_4'\cos 2k)\sin^2 k
                      -J_X(0)\cos^2 k]^2+
        8J_X(2k)^2\cos^2 k) \Bigg\} ^{1/2}
   \end{align}
\end{widetext}
   \end{subequations}
In both cuts, 
where $J_F(q)\equiv J_2+J_2'+2J_4\cos q+2J_4'\cos q$ is the effective ferromagnetic coupling 
and $J_X(q)=J_1+2J_3\cos q$ is the effective cross-sublattice coupling. 

\SAVE{These cuts included since minima tend to occur
along lines of high symmetry, and if the bifurcation occurs along this cut,
it simplifies the math (cubic eqns) dramatically.}

\subsection{State Perturbation (chain lattice example)} 
\label{sec:pert-state}

The major advantage of applying perturbation theory to a parametrization
of the lattice's spins, rather than the LT matrix, is that it can accomodate a more generic ground state. This is not say that it avoids the disconnect between the state being perturbed and the actual ground state, if the perturbed state does not encompass the true ground state then the disconnect is required.
However, this method does allow us to consider states composed of multiple LT wave-vectors.

In the octahedral lattice, most of the ground state types we encountered -- even
the noncollinear ones -- are essentially built using a single ``star'' of 
symmetry-related ordering wave-vectors.  
\SAVE{(The exception was the asymmetric conic, but that is only stabilized
by higher-order couplings, see Sec.~\ref{sec:phase-octahedral}.)}
Therefore, for the octahedral lattice, the actual phase diagram mostly
reflects the LT phase diagram and it is preferable to find phase boundaries
via the LT perturbation method illustrated above in Sec.~\ref{sec:pert-LT}.
For the chain lattice, however, the ground state is typically characterized by 
several wave-vectors. 

(as noted near the beginning of Sec.~\ref{sec:conic})

Such multi-LT wave-vectors are all parametrized by some form 
of ``conic spirals,'' a mix of a planar spiral using the wavevector 
$(q)$ and a deviation along the wavevector $(k_2)$
where $k_2$ is either integer or half-integer 
(since the chain lattice is 1D, the wave-vectors are as well). 
Within a range of parameter space $j_1$ through $j_4/j_4'$ ($j_5$
and higher all 0), these conic spirals are the most general form of
ground state, making the problem of finding the ground state amenable
to variational methods. 

As an illustration of determining the ground
state by variational methods, we consider the problem of finding the
phase boundaries for the {}``alternating conic'' $k_2=1/2$.
The spin configurations in the lattice are parametrized by Eq.\eqr{eq:altconic}.
Without loss of generality, we take the odd sites to be the ones with planar spins.
A symmetry relates the states \eqr{eq:altconic}
to the other family of alternating conic configurations
in which even and odd sites have swapped roles, if
one also swaps in parameter space 
$(j_2,j_4)\leftrightarrow(j_2',j_4')$.

Up to interaction $j_4$, the energy per unit cell is
    \begin{multline}
     E  =  -2(j_1\cos\psi+j_3\cos 3\psi)\cos\alpha\\
        -2(j_2\cos^2\psi-j_4\sin^2 2\psi)\cos^2\alpha\\
        +j_2-j_4-j_2'\cos 2\psi-j_4'\cos 4\psi 
   \label{eq:E-alt-conic}
    \end{multline}
Setting $\partial E/\partial (\cos \alpha)=0$ to minimize \eqr{eq:E-alt-conic},
we see that the optimal angle $\alpha^{*}$, is given by
   \begin{equation}
        \cos\alpha^*=-\frac{1}{2\cos\psi}
            \frac{j_1+j_3(4\cos^2 \psi-3)}
                 {j_2-4j_4\sin^2 \psi}
   \label{eq:alpha-alt-conic}
   \end{equation}
Notice the symmetry under reversing the signs of $j_1$ and
$j_3$ and $\alpha \leftrightarrow \pi -\alpha$.
Of course, a necessary condition is that the r.h.s. of
\eqr{eq:alpha-alt-conic} lies in $(0,+1)$ (recall that $0\le\alpha\le\pi/2$ by definition), otherwise
$\alpha^*$ is pinned to 0 or $\pi/2$, which would be a
planar spiral (or possibly a collinear state, depending
on the value of $\psi$).

Plugging \eqr{eq:alpha-alt-conic} into Eq.~\eqr{eq:E-alt-conic}
leaves $\psi =q/2$ as the only variational parameter:
   \label{eq:E-alt-conic-psi}
   \begin{eqnarray}
E & = & \frac{2j_3^2}{j_4}
       \frac{\Big(\frac{j_1-3j_3}{4j_3} + \cos^2\psi\Big)^2}
       {\frac{j_2-4j_4}{4j_4}+\cos^2\psi}
           -j_2'\cos 2\psi- \nonumber \\
   && \;\;\;\; - j_4'\cos 4\psi-j_2-j_4
     \end{eqnarray}
If we drop the secondary couplings equation \eqr{eq:E-alt-conic-psi}
reduces to
   \begin{equation}
   \label{E-alt-conic-simple}
        E  = \frac{1}{2}\frac{j_1^2}{j_2}-j_2'\cos 2\psi
   \end{equation}
Since \eqr{E-alt-conic-simple}
is linear in $\cos 2\psi$, its ground state
given by $\psi_m=m\pi/2,$ with $m$ integer.
If $m$ is odd, then Eq.\ref{E-alt-conic-simple} breaks down, since Eq.\eqr{eq:alpha-alt-conic} is singular. What actually occurs here is that the sublattices have adopted an antiferromagnetic structure, decoupling the sublattices. Because the sublattices are decoupled, $\alpha$ is arbitrary (representing the freedom the decoupled sublattices to rotate relative to each other). The energy for such a configuration is $E=j_2+j_2'$.
Conversely, if $m$ is even, then the ground state is a splayed state (assuming that $|j_1|<|2j_2|$)
That means a commensurate, {\it planar} state
(using the $xz$ plane, or collinear if $\alpha^*$
is trivial).  Thus, with only primary couplings, 
even though the two sublattices are inequivalent,
we {\it cannot} obtain a non-coplanar spiral.

When we turn on (not too large) secondary couplings, 
those commensurate states will remain stable out to
the critical coupling, at which the optimal wavevector
bifurcates.  Therefore we expand \eqr{eq:E-alt-conic-psi}
in powers of $\delta \equiv \psi-\psi_m$ 
about both kinds of stationary point~\cite{FN-conic-pert}
to find:
   \begin{subequations}
   	\begin{equation}
   		\begin{split}
       E(\delta)   =&  
          \frac{(j_1+j_3)^2}{2j_2}
          -j_2'-j_4'-j_2-j_4+ \\
      & [2\frac{(j_1+j_3)^2}{j_2^2}j_4- 
       4[j_1-j_3] \frac{j_1}{j_2}+2j_2'+8j_4']
        \delta^2
		   \end{split}
   	\end{equation}
	   \begin{equation}
   		 \begin{split}
       E(\frac{\pi}{2} + \delta)  =&  
           \frac{1}{2}\frac{[j_1-3j_3]^2}{j_2-4j_4}
           +j_2'-j_4'-j_2-j_4+ \\
     &     \Bigg(4\frac{j_1-3j_3}{j_2-4j_4} j_3+
            2[\frac{j_1-3j_3}{j_2-4j_4}]^2 j_4-
            2j_2'+8j_4'\Bigg)
           \delta^2
    		\end{split}
	    \end{equation}
    \end{subequations}
The commensurate state becomes unstable to $\delta\neq 0$
when the coefficient of the quadratic
term goes negative, so the conditions to induce instability are: 
  \begin{subequations}
  \label{eq:E-alt-conic-unstable}
     \begin{align}
&E(\delta): 0  \ge & 
     \frac{(j_1+j_3)^2}{j_2^2}j_4
      -2\frac{[j_1-j_3]} {j_2}j_3
        +j_2'+4j_4'  \\
& E_{\frac{\pi}{2}}: 0  \ge & 
     [\frac{j_1-3j_3}{j_2-4j_4}]^2j_4
        +2\frac{j_1-3j_3} {j_2-4j_4}j_3
         -j_2'+4j_4'
         \end{align}
   \end{subequations}
Combined with the requirement that $|\cos\alpha|<1$
in   \eqr{eq:alpha-alt-conic},
Eqs.~\eqr{eq:E-alt-conic-unstable}
give the minimum necessary conditions for the existance of an
alternating conic state.


\begin{thebibliography}{99}
\bibitem{FN-strucfactor}
This is not quite the magnetic structure factor,
since that includes interferences between different
sublattices.
\bibitem{FN-higherD}
Occasionally the end-to-end plot is illuminating
even in higher dimensions, the sequence of spins
being defined by a raster scan that goes down
rows -- limited to one of the three sublattices
-- running in one of the coordinate directions.)
For example, the double-twist state (Sec.~\ref{sec:doubletwist})
was discovered from common-origin and end-to-end plots.
\bibitem{FN-asymm-conic}
We did not include the asymmetric conic state in table \ref{tab:O-States}
since we will see that it is only found on the phase boundary between two other states. 
For the octahedral lattice, then, it appears only as a randomly chosen
member of a family of highly degenerate states
(using the classification of Sec.~\ref{sec:bridging}),
rather than a genuine ground state.
\bibitem{FN-spurious-AC}
Around the tricritical-type points where the transition
switches from first-order to degenerate, there are alternating conic 
states only slightly higher in energy than the helimagnetic
states, so that a sliver of alternating conic may be spuriously
identified here if numerical results are not checked analytically.
\bibitem{FN-conic-pert}
There are just two kinds of stationary point since
\eqr{E-alt-conic-simple} has period $\pi$.
\end{thebibliography}
\end{document}